\documentclass[12pt]{article}
\pdfoutput=1
\usepackage{putex}
\usepackage{graphicx}
\usepackage{epstopdf}
\usepackage{enumerate}
\usepackage{cite}
\usepackage{tensor}
\usepackage{slashed}

\numberwithin{equation}{section}

\newcommand{\HH}{\mathbb{H}}

\newcommand{\abs}[1]{\left\lvert #1 \right\rvert}

\newcommand {\be} {\begin {equation}}
\newcommand {\ee} {\end {equation}}

\newcommand {\bes} {\begin {equation*}}
\newcommand {\ees} {\end {equation*}}

\newcommand{\es}[2] {\begin{equation} \label{#1} \begin{split} #2 \end{split} \end{equation}}

\newcommand{\Z}{\mathbb{Z}}

\newcommand{\R}{\mathbb{R}}
\newcommand{\C}{\mathbb{C}}

\newcommand{\beq}{\begin{equation}}
\newcommand{\eeq}{\end{equation}}

\usepackage{cite,dsfont}
\usepackage{amsfonts}
\usepackage{amssymb}
\usepackage{amsmath}
\usepackage{graphicx}
\usepackage{slashed}
\usepackage{setspace}

\usepackage{color}

\newcommand{\cn}{{\cal N}}

\newcommand{\reef}[1]{(\ref{#1})}

\newcommand{\co}{{\cal O}}

\def\be{\begin{equation}}
\def\ee{\end{equation}}
\def\bea{\begin{eqnarray}}
\def\eea{\end{eqnarray}}
\def\ba{\begin{array}}
\def\ea{\end{array}}
\def\bd{\begin{displaymath}}
\def\ed{\end{displaymath}}

\def\tr{{\rm tr}}

\def\a{\alpha}
\def\b{\beta}

\def\d{\delta}
\def\e{\epsilon}           % Also, \varepsilon
\def\g{\gamma}
\def\h{\eta}

\def\l{\lambda}
\def\m{\mu}
\def\n{\nu}
  
\def\r{\rho}                                     %     \varrho
\def\s{\sigma}                                   %     \varsigma

\def\L{\Lambda}

\def\pa{\partial}                              % curly d
\def\>{\rangle} %right angle
\def\<{\langle} %left angle
\def\Dsl{D \hskip-.6em \raise1pt\hbox{$ / $ } }
\def\to{\rightarrow}
\def\pa{\partial}

\def\lab{\label}

\newcommand{\eps}{\epsilon}

\def\tz{{\tilde{z}}}

\def\bb{\bar{\beta}}
\def\bz{\bar{z}}
\def\bg{\bar{\gamma}}
\def\bW{\bar{W}}
\def\ba{\bar{a}}

\begin{document}

\preprint{SU-ITP-13/01\\
MIT-CTP-4443}

\institution{MITCTP}{Center for Theoretical Physics, Massachusetts Institute of Technology, Cambridge, MA 02139, USA}
\institution{MIT}{Department of Mathematics, Massachusetts Institute of Technology, Cambridge, MA 02139, USA}
\institution{Stanford}{Stanford Institute for Theoretical Physics, Department of Physics, Stanford University,\cr Stanford, CA 94305, USA}
\institution{Princeton}{Joseph Henry Laboratories, Princeton University, Princeton, NJ 08544, USA}

\title{The Holography of $F$-maximization}

\authors{Daniel Z.~Freedman\worksat{\MITCTP, \MIT, \Stanford} and Silviu S.~Pufu\worksat{\MITCTP, \Princeton}}

\abstract{
We find new supersymmetric backgrounds of ${\cal N} = 8$ gauged supergravity in four Euclidean dimensions that are dual to deformations of ABJM theory on $S^3$.  The deformations encode the most general choice  of $U(1)_R$ symmetry used to define the theory on $S^3$.  We work within an ${\cal N} = 2$ truncation of the ${\cal N} = 8$ supergravity theory obtained via a group theory argument.  We find perfect agreement between the $S^3$ free energy computed from our supergravity backgrounds and the previous field theory computations of the same quantity based on supersymmetric localization and matrix model techniques.
 }

\date{January, 2014}

\maketitle

\tableofcontents

\section{Introduction}

The past few years have seen much progress in our understanding of supersymmetric quantum field theories in $2+1$ space-time dimensions.  From the perspective of string theory and the AdS/CFT duality \cite{Witten:1998qj,Gubser:1998bc,Maldacena:1997re}, a significant result was the discovery of $(2+1)$-dimensional superconformal field theories (SCFTs) on $N$ coincident M2-branes placed at the tip of various Calabi-Yau cones.  The simplest example is the case where the Calabi-Yau cone is $\C^4 / \Z_k$ and the dual SCFT is the ABJM theory, a $U(N)_k \times U(N)_{-k}$  
Chern-Simons-matter gauge theory with ${\cal N} = 6$ supersymmetry \cite{Aharony:2008ug} (for a review, see for instance \cite{Klebanov:2009sg};  for earlier work, see \cite{Nishino:1991sr, Gates:1991qn, Schwarz:2004yj, Bagger:2006sk, Bagger:2007jr, Bagger:2007vi, Gustavsson:2007vu}).  Independent of string theory, there have also been new exact results such as computations of partition functions of supersymmetric theories on curved manifolds \cite{Kapustin:2009kz, Hama:2011ea} that use the technique of supersymmetric localization introduced in \cite{Witten:1988ze, Pestun:2007rz}, as well as a procedure called $F$-maximization \cite{Jafferis:2010un,Jafferis:2011zi, Closset:2012vg}.
$F$-maximization states that in ${\cal N} = 2$ SCFTs, the $U(1)_R$ symmetry that appears in the superconformal algebra is precisely the one that maximizes $F = -\log \abs{Z_{S^3}}$, $Z_{S^3}$ being the partition function on $S^3$, over the set of all possible $U(1)$ R-symmetries.   The present paper searches for an AdS/CFT interpretation of $F$-maximization, where it focuses on the particular case of ABJM theory and its supergravity dual.

In ${\cal N} = 2$ notation, the matter content of the $U(N)_k \times U(N)_{-k}$ ABJM theory consists of two bifundamental chiral multiplets $Z^a$, $a=1, 2$, transforming in the representation $(\bar {\bf N}, {\bf N})$ of the gauge group and two bifundamental chiral multiplets $W_b$, $b=1, 2$, transforming in the conjugate representation $({\bf N}, \bar {\bf N})$.  The most general choice of a $U(1)_R$ symmetry group within the global symmetry group of ABJM theory involves three parameters.  These parameters
determine an assignment of R-charges\footnote{This description is somewhat imprecise because the matter chiral multiplets are not gauge invariant, so the $R[Z^a]$ and $R[W_b]$ may not be gauge invariant observables.  However, one can construct gauge invariant observables by combining the bifundamental chiral multiplets with certain monopole operators.  Under the convention that the monopole and antimonopole operators have the same R-charge, the $R[Z^a]$ and $R[W_b]$ become well-defined observable quantities.  For more details, see \cite{Jafferis:2011zi}. \label{MonopoleFootnote}}  
$R[Z^a]$ and $R[W_b]$ to the four bifundamental chiral multiplets
such that the constraint
\es{Constraint}{
   R[Z^1] + R[Z^2] + R[W_1] + R[W_2] = 2 
 }
coming from the requirement that the quartic superpotential $W \propto \tr \left( \epsilon_{ab} \epsilon^{cd} Z^a W_c Z^b W_d \right)$ has R-charge two
is satisfied.  For a general choice of the  $U(1)_R$ charges $R[Z^a]$ and $R[W_b]$, one can define a deformed ABJM theory on $S^3$ by specifying couplings to curvature that preserve only an $OSp(2 |2) \times SU(2)$ super-algebra that contains $U(1)_R \subset OSp(2 |2)$.  The reason why the $S^3$ Lagrangian depends on the $U(1)_R$ charges is that the minimal algebra on $S^3$ that includes four supercharges must also include a $U(1)_R$ symmetry, in stark contrast with the flat space situation.

For a general choice of $U(1)_R$, the theory on $S^3$ will not be conformal.   The theory is conformal only when $R[Z^a] = R[W_b] = 1/2$, as can be deduced from the embedding of $OSp(2|2) \times SU(2)$ into the superconformal algebra $OSp(6|2, 2)$ corresponding to ${\cal N} = 6$ superconformal symmetry.  (When $k=1, 2$ the superconformal algebra is enhanced further to $OSp(8 | 2, 2)$ corresponding to ${\cal N} =8$ superconformal symmetry.)  If one did not know that the superconformal ABJM theory had more than ${\cal N} = 2$ supersymmetry, one would have to resort to $F$-maximization to determine which $U(1)_R$ charges correspond to the superconformal theory.  Using matrix model techniques that build on the supersymmetric localization results of \cite{Kapustin:2009kz, Jafferis:2010un}, it was found to leading order in $N$ that \cite{Jafferis:2011zi}
 \es{FreeEnergy}{
   F = \frac{\sqrt{2} \pi N^{3/2}}{3} 4 \sqrt{R[Z^1] R[Z^2] R[W_1] R[W_2]} \,,
 }
when the Chern-Simons level is $k=1$, which is the case we will focus on from here on.  Under the constraint \eqref{Constraint}, this expression is maximized for the free-field values $R[Z^a] = R[W_b] = 1/2$, which correspond to the superconformal theory.  For any other choice of the $U(1)_R$ charges, the $S^3$ theory is not conformal, and we can think of it as a relevant deformation of the superconformal one.

In the large $N$ approximation, the M-theory dual of the superconformal ABJM theory with $k=1$ is given by the $AdS_4 \times S^7$ vacuum of eleven-dimensional supergravity.\footnote{For arbitrary $k$ the supergravity dual is $AdS_4 \times S^7/\Z_k$, and the eleven-dimensional supergravity approximation is valid as long as $N \gg k^5$ \cite{Aharony:2008ug}.}  Instead of working in eleven dimensions, we work within a consistent truncation of ${\cal N} = 8$ gauged supergravity in four dimensions \cite{deWit:1982ig} and find a three-parameter family of classical solutions dual to the general deformation of ABJM theory discussed above.  The acid test of our work is the agreement between the $S^3$ free energy $F$ calculated from supergravity with the field theory expression \eqref{FreeEnergy}.   In particular,  when the free energy is maximized, the bulk geometry becomes (Euclidean) $AdS_4$ (more correctly, the hyperbolic space $\HH^4$), with the expected superconformal symmetry.

Our classical supergravity solutions describe holographic RG flows that start from the superconformal ABJM theory in the UV, but never reach a true IR limit, because on $S^3$ one cannot probe distances larger than the radius of the sphere.   These solutions are nevertheless smooth:   the three-sphere shrinks to zero size smoothly at some value of the holographic radial coordinate.  (See \cite{Martelli:2011fw, Martelli:2011fu, Martelli:2012sz} for other supergravity solutions dual to field theories on compact spaces, where a similar phenomenon occurs.)  In some sense, the existence of a largest length scale on $S^3$ prevents the appearance of a singularity in the bulk.

The $AdS_4$ solution of ${\cal N} =8$ gauged supergravity is invariant under the $OSp(8|4)$ superalgebra, 
the same algebra as the dual ABJM theory at Chern-Simons level $k=1$.  The classical solutions we find are Euclidean solutions that break  this large symmetry to the Euclidean superalgebra $OSp(2|2)\times SU(2)$.  In view of the discussion of $F$-maximization above, it suffices to look for a consistent $\cn=2$ truncation of the theory of \cite{deWit:1982ig}.  
Our solutions involve only fields in the $\cn=2$ truncation.  We use symmetries to guide us to a truncated theory that contains the $\cn=2$ gravity multiplet plus three $U(1)$ gauge multiplets.  The asymptotic values of the three complex scalars of the gauge multiplets determine a choice of R-charges in the perturbed ABJM theory on the boundary.  Our supergravity solutions are extrema of the bosonic Euclidean action
 \es{BulkAction}{
   S = \frac{1}{8 \pi G_4} \int d^4 x\, \sqrt{g} \left[ -\frac 12 R +\sum_{\alpha = 1}^3 \frac{\abs{\partial_\mu z^\alpha}^2 }{\left(1 - \abs{z^\alpha}^2 \right)^2} 
   + \frac{1}{L^2} \left( 3 - \sum_{\alpha = 1}^3 \frac{2}{1 - \abs{z^\alpha}^2} \right) \right] \,,
 }
 where $G_4$ is the Newton constant in four dimensions, and $L$ is a constant chosen such that one of the extrema of \eqref{BulkAction} is $\HH^4$ of radius $L$.  The action for the scalar fields is a sigma model with $\HH^2 \times \HH^2 \times \HH^2$ target space.  When $L=\infty$, it reduces to the well known $stu$-model \cite{Duff:1995sm, Behrndt:1996hu}.  The action \reef{BulkAction} may be completed to a gauged $\cn =2$ supergravity model by adding appropriate gauge field\footnote{The gauge fields were omitted from \eqref{BulkAction} because they vanish in the classical solutions needed to describe the duals of the deformed ABJM theory.} and fermion terms.
In this form it contains a  $SO(2)_R \times SO(2)^3$  residue of the
$SO(8)_R$ symmetry of the parent $\cn=8$  theory.  The $SO(2)_R$ rotates the two gravitino fields and will be identified with the $U(1)_R$ of the deformed ABJM theory.

There are four main subtleties related to the action \eqref{BulkAction}, its extrema, and their field theory interpretation.  The first subtlety concerns the group theory needed to derive \eqref{BulkAction} as a consistent truncation of ${\cal N} = 8$ gauged supergravity.  The $70$ real scalar fields of the ${\cal N} = 8$ theory transform in the ${\bf 35}_v$ and ${\bf 35}_c$ representations of the $SO(8)_R$ symmetry, and they are customarily written as self-dual and anti-self dual fourth-rank antisymmetric products of the ${\bf 8}_s$ representation \cite{deWit:1982ig}.  The field theory deformations we are interested in can be written as Lagrangian deformations in a formalism where only an $SU(4)_R$ subgroup of $SO(8)_R$ is manifest.  Even if the whole $SO(8)_R$ were manifest, it would be more natural to write down these deformations as states in ${\bf 35}_v$ and ${\bf 35}_c$ represented as rank-two traceless symmetric products of ${\bf 8}_v$ and ${\bf 8}_c$.  Making the connection between the two ways of representing ${\bf 35}_v$ and ${\bf 35}_c$ and examining their decomposition under $SU(4)_R$ (and further subgroups thereof) requires a tedious group theory analysis.  This analysis is needed to establish an explicit correspondence between the three $z^\alpha$ appearing in \eqref{BulkAction} and their dual field theory operators written in terms of the bifundamental fields of the ABJM theory.
 
The second subtlety involves the Euclidean continuation of Lorentzian supergravity.  It is well known that spinors $\psi$ and $\psi^\dagger$ that are related by complex conjugation in a Lorentzian theory become independent in its Euclidean continuation.  The reason is that the isometry group of flat Euclidean space factors as $SO(4) = SU(2)\times SU(2)$, and group elements are described by the pair $(U,V)$ of $2 \times 2$ unitary matrices.  Since $\psi \to U\psi$, but $\psi^\dagger \to  \psi^\dagger V^{-1}$, we must allow $\psi$ and $\psi^\dagger$ to be independent.   (This contrasts with the Lorentz group in which the $(1/2,0)$ and $(0,1/2)$ are conjugate representations.)  Since fermions and bosons are linked by SUSY transformation rules, we must also allow formally conjugate boson fields to be independent. 
 In particular, in the Euclidean action \eqref{BulkAction} one should treat as independent the complex fields $z^\alpha$ and their would-be Lorentzian conjugates $\bar z^\alpha$. (Later we denote the conjugate scalars by $\tilde z^\alpha$ to emphasize that they are not related to the $z^\alpha$).

The third subtlety is related to holographic renormalization.  One must choose a renormalization scheme that is consistent with supersymmetry.  To find such a scheme one should understand how supersymmetry transformations act asymptotically on the fields $z^\alpha$ and $\tilde z^\alpha$.  A similar analysis was performed in \cite{Amsel:2008iz} in Lorentzian signature and provides a different way of understanding that, up to a chiral rotation, supersymmetry requires the real and imaginary parts of $z^\alpha$ to be quantized with opposite boundary conditions \cite{Breitenlohner:1982jf}.  In our Euclidean setup it is $z^\alpha - \tilde z^\alpha$ that should be quantized using regular boundary conditions and $z^\alpha + \tilde z^\alpha$ that should be quantized using the alternate boundary conditions described in \cite{Klebanov:1999tb}.

The fourth subtlety is related to the last point just mentioned.  It is not the renormalized bulk on-shell action that should be identified with the boundary free energy $F$, but instead its Legendre transform with respect to the leading asymptotic behavior of $z^\alpha + \tilde z^\alpha$.  Such a Legendre transform was introduced in \cite{Klebanov:1999tb}, where it was explained that the Legendre transform is necessary for obtaining the correct correlation functions in the field theory whose gravity dual contains a scalar with alternate boundary conditions. 

In the following sections we provide more detailed information.  In section \ref{FIELDTHEORY} we describe the field theory setup more carefully.  An important  viewpoint,  advocated in \cite{Festuccia:2011ws, Closset:2012vg},
is that  one should think of the three-parameter family of R-charge assignments \eqref{Constraint} as complexified ${\cal N} = 2$-preserving real mass deformations of ABJM theory with purely imaginary mass parameters.  In three dimensions, a real mass deformation arises not from a superpotential but from the coupling of a $U(1)$ current multiplet to a background ${\cal N} = 2$ vector multiplet with SUSY-preserving expectation values for the scalar fields of the vector multiplet.  In section~\ref{TRUNCATION} we derive an ${\cal N} = 2$ consistent truncation of ${\cal N} = 8$ gauged supergravity, which we further describe in an ${\cal N} = 1$ formulation whose bosonic Lagrangian is \eqref{BulkAction}.  In section~\ref{CONTINUATION} we describe the analytic continuation to Euclidean signature and provide the Euclidean supersymmetry transformation rules.  In section~\ref{BPS} we derive and solve the BPS equations that follow from these transformation rules.  Lastly, in section~\ref{HOLOGRAPHY} we perform holographic renormalization and give the field theory interpretation of our solutions.  We will be able to reproduce \eqref{FreeEnergy} from a gravity calculation.  Many of the details of our computations are relegated to the Appendices.

\section{Field theory setup}
\label{FIELDTHEORY}

\subsection{${\cal N} =2$ deformations of ABJM theory}

As mentioned in the introduction, ABJM theory with gauge group $U(N) \times U(N)$ and Chern-Simons levels $(k,-k)$ for the two gauge group factors has ${\cal N} = 6$ supersymmetry and global $SU(4)_R \times U(1)_b$ symmetry group.  When $k = 1,2$, this symmetry is enhanced to $SO(8)_R$, and hence the theory has ${\cal N} =8$ supersymmetry.  In the ${\cal N} = 2$ formulation presented in \cite{Aharony:2008ug}, the symmetry group that acts on the ${\cal N} = 2$ super-fields is a $U(1)_R \times SU(2) \times SU(2) \times U(1)_b$ subgroup of $SU(4)_R \times U(1)_b$.  The $SU(4)_R \times U(1)_b$ symmetry becomes visible only when the Lagrangian is written in terms of the ${\cal N} = 2$ super-field components \cite{Aharony:2008ug, Benna:2008zy}.

In ${\cal N} = 2$ notation, the field content of ABJM theory consists of the two $U(N)$ vector multiplets and four chiral multiplets that transform in bifundamental representations of the gauge group.  An off-shell ${\cal N} = 2$ vector multiplet $(A_i, \sigma, \lambda, D)$ in 3d consists of a vector field $A_i$, a complex fermion $\lambda$, and two real scalars $\sigma$ and $D$.\footnote{In Euclidean signature, the gauge field $A_i$ and the scalars $\sigma$ and $D$ are allowed to take complex values, and $\lambda$ should be treated as independent from its complex conjugate.  Similarly, the chiral multiplet fields $(Z, \chi, F)$ introduced later on should be considered as independent from their complex conjugates.}  We denote the two $U(N)$ vector multiplets of ABJM theory by $(A_i, \sigma, \lambda, D)$ and $(\tilde A_i, \tilde \sigma, \tilde \lambda, \tilde D)$.  An ${\cal N} = 2$ chiral multiplet $(Z, \chi, F)$ consists of a complex scalar $Z$, a complex fermion $\chi$, and an auxiliary complex scalar $F$.  In ABJM theory we have two chiral multiplets $(Z^a, \chi^a, F^a)$, $a = 1, 2$, that transform in the $({\bf \bar N}, {\bf N})$ representation of the $U(N) \times U(N)$ gauge group, and two chiral multiplets $(W_a, \eta_a, G_a)$, $a = 1, 2$, that transform in the conjugate representation $({\bf N}, {\bf \bar N})$.  The gauge and global $U(1)_R \times SU(2) \times SU(2) \times U(1)_b$ charges of all the fields are summarized in Table~\ref{ABJMFieldsTable}.  The salient features are as follows.
\begin{table}[htdp]
 \begin{center}
 \begin{tabular}{c|c|c|c|c}
  field & $U(N) \times U(N)$ & $SU(2) \times SU(2)$ & $U(1)_R$ & $U(1)_b$ \\
  \hline
  \hline
  $(A_\mu, \sigma, \lambda, D)$ & $(\text{adj}, {\bf 1})$ & $({\bf 1}, {\bf 1})$ & $(0, 0, 1, 0)$ & $0$ \\
  $(\tilde A_\mu, \tilde \sigma, \tilde \lambda, \tilde D)$ & $({\bf 1}, \text{adj})$ & $({\bf 1}, {\bf 1})$ & $(0, 0, 1, 0)$ & $0$ \\  
  $(Z^a, \chi^a, F^a)$ & $({\bf \bar N}, {\bf N})$ & $({\bf 2}, {\bf 1})$ & $(1/2, -1/2, -3/2)$ & $1$ \\
  $(W_a, \eta_a, G_a)$ & $({\bf N}, {\bf \bar N})$ & $({\bf 1}, {\bf \bar 2})$ & $(1/2, -1/2, -3/2)$ & $-1$ \\
 \end{tabular}
 \caption{The fields of ABJM theory and their gauge and global charges.}
  \label{ABJMFieldsTable} 
 \end{center}
\end{table}
The $U(1)_R$ charges are those of free fields.   The first $SU(2)$ factor rotates the first pair of chiral multiplets, and the second $SU(2)$ factor rotates the second pair.   The $U(1)_b$ symmetry is generated by the topological current $* \tr (F + \tilde F)$, where $F$ and $\tilde F$ are the field strengths of the two $U(N)$ gauge fields.   The gauge-invariant objects with charge $n$ contain monopole operators $T^{(n)}$ that turn on $n$ of units of $* \tr (F + \tilde F)$ flux through a two-sphere surrounding the insertion point.  When $k=1$, $T^{(1)}$ transforms in $({\bf N}, {\bf \bar N})$ and $T^{(-1)}$ transforms in $({\bf \bar N}, {\bf N})$, and we can construct gauge-invariant operators such as $\tr (T^{(1)} Z^a)$ and $\tr (T^{(-1)} W_b)$.  The global $U(1)$ charges we listed in Table~\ref{ABJMFieldsTable} are in the convention that the monopole $T^{(1)}$ and the anti-monopole $T^{(-1)}$ have equal charges---for more details, see \cite{Jafferis:2011zi}.

The Lagrangian of the superconformal ABJM theory consists of Chern-Simons kinetic terms for the two ${\cal N} = 2$ vector multiplets, standard kinetic terms for the chiral multiplets $Z^a$ and $W_b$, and superpotential interaction terms coming from a superpotential of the form 
 \es{SuperpotFieldThy}{
   W \propto \tr \left( \epsilon_{ab} \epsilon^{cd} Z^a W_c Z^b W_d \right) \,.
 }  
The superconformal R-charge assignment in Table~\ref{ABJMFieldsTable} follows from the fact that $W$ should have R-charge two and that the $SO(8)_R$ symmetry mixes together the $Z^a$ and $W_b$. 

In this paper we break the global symmetry group of ABJM theory to its maximal Abelian subgroup by considering an R-charge assignment different from that in Table~\ref{ABJMFieldsTable}.   The most general R-charge assignment consistent with the marginality of the superpotential \eqref{SuperpotFieldThy} can be taken to be
 \es{RGeneral}{
  R[Z^1] &= \frac 12 + \delta_1 + \delta_2 + \delta_3 \,, \qquad
    R[W_1] = \frac 12 -\delta_1 + \delta_2 - \delta_3 \,, \\
    R[Z^2] &= \frac 12 + \delta_1 - \delta_2 - \delta_3 \,, \qquad
  R[W_2] = \frac 12 -\delta_1 - \delta_2 + \delta_3 \,,
 }
where $\delta_\alpha$ are three parameters.   One can think of \eqref{RGeneral} as a mixing of the canonical R-symmetry from Table~\ref{ABJMFieldsTable} with the diagonal $U(1) \times U(1) \times U(1)_b$ subgroup of the $SU(2) \times SU(2) \times U(1)_b$ flavor symmetry.  The $U(1)_R$ symmetry with charges \eqref{RGeneral} is still a symmetry of ABJM theory on $\R^{2, 1}$, but this $U(1)_R$ symmetry is not the one that appears in the ${\cal N} = 2$ superconformal algebra $OSp(2 | 4) \subset OSp(8 | 4)$.

As explained in \cite{Jafferis:2010un, Closset:2012vg,Festuccia:2011ws}, given a $U(1)_R$ symmetry (which in general can be taken to be a linear combination of a canonical $U(1)_R$ symmetry and other flavor $U(1)$ symmetries, as in \eqref{RGeneral}) of an ${\cal N} = 2$ theory on $\R^{2, 1}$, one can construct a theory on $S^3$ that is invariant under $OSp(2 | 2)_r \times SU(2)_\ell$, where the bosonic part of $OSp(2 | 2)_r$ is $U(1)_R \times SU(2)_r$ and $SU(2)_\ell \times SU(2)_r \cong SO(4)$ is the isometry group of $S^3$.\footnote{Similarly, one can preserve $OSp(2 | 2)_\ell \times SU(2)_r$ where the bosonic subgroup of $OSp(2 | 2)_\ell$ is $U(1)_R \times SU(2)_\ell$.  The two choices are related by formally sending $a \to -a$ in all the formulas presented in this section. \label{ReflectionFootnote}}  Such a construction was performed in \cite{Jafferis:2010un} at the level of a microscopic Lagrangian, and in \cite{Closset:2012vg,Festuccia:2011ws} more abstractly by coupling the flat space theory to a set of background fields.  Following the approach in \cite{Jafferis:2010un}, which we will explain shortly in section~\ref{VECTOR}, the $OSp(2 | 2)_r \times SU(2)_\ell$-invariant Lagrangian of ABJM theory on $S^3$ with the R-charge assignment \eqref{RGeneral} is 
 \es{LagABJMDeformed}{
  {\cal L} &= {\cal L}_\text{SCFT} + \sum_{b=1}^2 \left( R[Z^b] - \frac 12 \right)
   \tr \left(\frac 1{a^2} Z_b^\dagger Z^b + \frac 1a \chi_b^\dagger \chi^b + \frac{2i}{a}  \left( \sigma Z^\dagger_b Z^b   - Z^\dagger_b \tilde \sigma Z^b \right)
   \right) \\
   &+ \sum_{b=1}^2 \left( R[W_b] - \frac 12 \right)
   \tr \left(\frac 1{a^2} W^{\dagger b}W_b + \frac 1a \eta^{\dagger b} \eta_b + \frac{2i}{a}  \left( \tilde \sigma W^{\dagger b} W_b   - W^{\dagger b} \sigma W_b \right)  \right) \\
   &- \frac{1}{a^2} \sum_{b=1}^2 \left( R[Z^b] - \frac 12 \right)^2 Z_b^\dagger Z^b
   - \frac{1}{a^2} \sum_{b=1}^2 \left( R[W_b] - \frac 12 \right)^2 W^{\dagger b} W_b \,.
 }
Here, ${\cal L}_\text{SCFT}$ is the Lagrangian of ABJM theory on $S^3$ with the canonical R-charge assignment corresponding to $\delta_\alpha = 0$.  The canonical R-charge assignment makes the theory superconformal, and ${\cal L}_\text{SCFT}$ can be obtained by conformally coupling the flat space ABJM Lagrangian to curvature.  In this paper we will find the supergravity dual of the theory with Lagrangian \eqref{LagABJMDeformed}.

Using \eqref{RGeneral}, \eqref{LagABJMDeformed} becomes 
 \es{LagPertABJM}{
  {\cal L} &= {\cal L}_\text{SCFT} +  \frac{1}{a^2} \left[ (\delta_1 - 2 \delta_2 \delta_3) {\cal O}_B^1
   + (\delta_2 - 2 \delta_1 \delta_3) {\cal O}_B^2 + (\delta_3 - 2 \delta_1 \delta_2) {\cal O}_B^3 \right] \\
  &+ \frac 1a \left(  \delta_1 {\cal O}_F^1 + \delta_2 {\cal O}_F^2 + \delta_3 {\cal O}_F^3 \right)
   - \frac{1}{a^2} (\delta_1^2 + \delta_2^2 + \delta_3^2) {\cal O}_S \,,
 }
where
 \es{OBDefs}{
   {\cal O}_B^1 &= \tr \left( Z_1^\dagger Z^1 + Z_2^\dagger Z^2 -W^{\dagger 1} W_1 - W^{\dagger 2} W_2 \right) \,, \\
   {\cal O}_B^2 &= \tr \left( Z_1^\dagger Z^1 - Z_2^\dagger Z^2 + W^{\dagger 1} W_1 - W^{\dagger 2} W_2 \right) \,, \\
   {\cal O}_B^3 &= \tr \left( Z_1^\dagger Z^1 - Z_2^\dagger Z^2 - W^{\dagger 1} W_1 + W^{\dagger 2} W_2 \right) \,, \\
   {\cal O}_S &= \tr \left( Z_1^\dagger Z^1 + Z_2^\dagger Z^2 + W^{\dagger 1} W_1 + W^{\dagger 2} W_2 \right) \,,
 }
and 
 \es{OFDefs}{ 
   {\cal O}_F^1 &= \tr \left( \chi^{\dagger 1} \chi_1 + \chi^{\dagger 2} \chi_2 - \eta_1^\dagger \eta^1 - \eta_2^\dagger \eta^2 \right) + \text{($\sigma$, $\tilde \sigma$ terms)} \,, \\
   {\cal O}_F^2 &= \tr \left( -\chi^{\dagger 1} \chi_1 + \chi^{\dagger 2} \chi_2 - \eta_1^\dagger \eta^1 + \eta_2^\dagger \eta^2 \right) + \text{($\sigma$, $\tilde \sigma$ terms)} \,, \\
   {\cal O}_F^3 &= \tr \left( -\chi^{\dagger 1} \chi_1 + \chi^{\dagger 2} \chi_2 + \eta_1^\dagger \eta^1 - \eta_2^\dagger \eta^2 \right) + \text{($\sigma$, $\tilde \sigma$ terms)} \,.
}
The terms involving $\sigma$ and $\tilde \sigma$ can be read off from \eqref{LagABJMDeformed} and were omitted in \eqref{OFDefs} for clarity.  The perturbation \eqref{LagPertABJM} breaks $SO(8)_R$ to its maximal Abelian subgroup $U(1)_R \times U(1) \times U(1) \times U(1)_b$.  Note that to leading order at small $\delta_\alpha$ the operator ${\cal O}_S$ is absent, and up to a factor of $a$ the coefficients of ${\cal O}_B^\alpha$ are equal to those of ${\cal O}_F^\alpha$.

Under the $SU(4)_R\times U(1)_b$ symmetry of the superconformal ABJM theory,
the scalars transform in the ${\bf 4}_1$ representation usually denoted by $Y^A = (Z^1, Z^2, W^{\dagger 1}, W^{\dagger 2})$.   It is clear from  \reef{OBDefs} that the $\co_B^\a$ are states in the adjoint ${\bf 15}_0$.  The  $\co_B^\a$ are the lowest components of the SUSY multiplet containing the conserved $SU(4)_R$ currents, so they have fixed scale dimension $\Delta =1$.  On the other hand, the operator $\co_S$ is an $SU(4)_R$ singlet, and its scale dimension is not protected from loop corrections.
There is no scalar dual to $\co_S$ in the supergravity theory we construct in Section~\ref{TRUNCATION}.  Nevertheless, the precise match of the free energy calculations in the gravity dual and in the deformed QFT indicates that the dynamical effects of $\co_S$  are included.

ABJM theory is invariant under a space-time parity symmetry that also exchanges the two gauge groups.  Under this symmetry ${\cal O}_B^\alpha$ are invariant, but ${\cal O}_F^\alpha$ change sign.  Therefore ${\cal O}_B^\alpha$ are scalar operators, while ${\cal O}_F^\alpha$ are pseudo-scalars.

\subsection{The deformed Lagrangian from coupling to background vector multiplets}
\label{VECTOR}

In this section we explain the construction of the deformed Lagrangian \eqref{LagABJMDeformed}.  Related  arguments appear in \cite{Jafferis:2010un, Festuccia:2011ws}.

An unusual feature of  the deformed Lagrangian \reef{LagABJMDeformed}  is that  the coefficients  $R[Z^b] - 1/2$, etc., which appear as coupling constants, actually denote an assignment  of R-charges to the elementary chiral operators of the ABJM theory.  A useful viewpoint is that \eqref{LagABJMDeformed} describes  ABJM theory on $S^3$ coupled to three background $U(1)$ vector multiplets, which take supersymmetry-preserving expectation values. 

Let us start with the simpler situation of a chiral multiplet  $(Z, \chi,F)$ 
interacting with an abelian vector multiplet  $(A_i, \s, \l,D)$
 and return to ABJM theory later.  In Appendix~\ref{SUSY3D} we outline a method to obtain this Euclidean theory
 via dimensional reduction from four dimensions  and to modify the supersymmetry transformations and Lagrangian when the theory is defined on $S^3$.  The main results are that the Lagrangian on the three-sphere of radius $a$ is
 \es{SchiralFree}{
 S_{1/2} &= \int d^3 x\, \sqrt{g} \Biggl(D^i Z^* D_i Z + \sigma^2 Z^* Z + i  \chi^\dagger \sigma^i D_i \chi 
   + i \chi^\dagger \sigma \chi  - F^* F \\
   &\qquad\qquad\qquad\qquad\qquad+ \lambda^T (i \sigma_2) Z^* \chi + \chi^\dagger (i \sigma_2) Z \lambda^* - D Z^* Z + \frac{3}{4a^2} Z^* Z \Biggr) \,,
 }
and that the supersymmetry algebra generated by $Q$ and $Q^\dagger$ is 
 \es{SUSYAlgebraMain}{
  \{ Q,  Q^\dagger \} = \sigma^i J_i  + i q \sigma + \frac{1}{a} R \,,
 } 
where $J_i$ is an $SU(2)_r$ isometry of $S^3$, $\sigma^i$ are the Pauli matrices, $q$ is the gauge charge, and $R$ is the $U(1)_R$ charge.  The gauge charges of $(Z, \chi, F)$ are $+1$, and the R-charges are $(1/2, -1/2, -3/2)$, as appropriate for free fields.  The  conjugate fields $(Z^*,\chi^\dag,F^*)$ have opposite gauge and R-charges.  The first and last terms in \eqref{SUSYAlgebraMain} correspond to the even generators of $OSp(2|2)_r$;  the middle term is just a gauge transformation.

As noticed in \cite{Jafferis:2010un}, one can also write down an $S^3$ Lagrangian that is invariant under a modified supersymmetry algebra generated by $Q'$ and $Q'^\dagger$ with
 \es{SUSYAlgebraDelta}{
  \{ Q',  Q'^\dagger \} = \sigma^i J_i  + i q \sigma + \frac{1}{a} R' \,,
 } 
such that the $U(1)_{R'}$ charges of a chiral multiplet are now $(\Delta, \Delta - 1, \Delta - 2)$ for some given $\Delta$.  The modified Lagrangian is
 \es{SchiralDelta}{
  S_\Delta = S_{1/2} + \int d^3 x\, \sqrt{g} \left[-\frac{1}{a^2} \left(\Delta - \frac 12 \right) \left( \Delta - \frac 32 \right) Z^* Z 
     + \frac {1}{a} \left( \Delta - \frac 12 \right) \left(\chi^\dagger \chi - \sigma Z^* Z \right)   \right] \,.
 }
This Lagrangian was obtained in \cite{Jafferis:2010un} by direct computation.  

A more conceptual way to derive \eqref{SchiralDelta} is \cite{Festuccia:2011ws} to notice that one can obtain \eqref{SUSYAlgebraDelta} from \eqref{SUSYAlgebraMain} by shifting 
 \es{GaugeShift}{
  (A_i, \sigma, \lambda, D) \to (A_i, \sigma, \lambda, D) + (A_i', \sigma', \lambda', D')
 }
and regarding $(A_i', \sigma', \lambda', D')$ as a background vector multiplet which is set to
 \es{BackgroundVector}{
  \sigma' = -i \frac{\Delta - 1/2}{a}\,, \qquad 
   D'= -\frac{\Delta - 1/2}{a^2} \,, \qquad A_i' = 0 \,, \qquad \lambda' = 0 \,.
 }
The background \eqref{BackgroundVector} is chosen so that it is invariant under supersymmetry.  Indeed, the supersymmetry variations of the bosonic fields $A_i'$, $\sigma'$, and $D'$ vanish automatically because they must be proportional to the fermions, which are set to zero.  The supersymmetry variation of $\lambda'$ is 
 \es{lambdaprimeSUSY}{
  \delta \lambda' = \left(\frac 12 \sigma^{ij} F_{ij}' + i \sigma^i \partial_i \sigma' + i D' - \frac 1a \sigma' \right) \epsilon \,,
 }
where $\epsilon$ is a left-invariant Killing spinor on $S^3$ (see \eqref{KspS3}), and it can be easily seen that it also vanishes in the background \eqref{BackgroundVector}.  Since the supersymmetry variations of $\sigma'$ and $D'$ vanish, we can treat these quantities as coupling constants in the Lagrangian.

Let's return now to the case of ABJM theory that we are interested in here.  The SCFT Lagrangian, obtained for free-field R-charge assignments, is that corresponding to the non-Abelian generalization of \eqref{SchiralFree} for each of the chiral multiplets $(Z^a, \chi^a, F^a)$ and $(W_a, \eta_a, G_a)$, as well as Chern-Simons kinetic terms for the vector multiplets and a superpotential interaction derived from \eqref{SuperpotFieldThy}.  The most general R-charge assignment \eqref{RGeneral} is obtained by coupling ABJM theory to three background vector multiplets $(A_i'^\alpha, \sigma'^\alpha, \lambda'^\alpha, D'^\alpha)$ and setting
 \es{ThreeBackground}{
   \sigma'^\alpha = -i \frac{\delta_\alpha}{a}\,, \qquad 
   D'^\alpha = -\frac{\delta_\alpha}{a^2} \,, \qquad A_i'^\alpha = 0 \,, \qquad \lambda'^\alpha = 0 \,.
 }
Just like \eqref{BackgroundVector}, this background also preserves supersymmetry.    To reproduce \eqref{RGeneral} we should take the charges of the fields $Z^1$, $Z^2$, $W_1$, $W_2$ (and of their SUSY partners) under the three background vector multiplets to be $(1, 1, -1, -1)$, $(1, -1, 1, -1)$, and $(1, -1, -1, 1)$.  The Lagrangian \eqref{LagABJMDeformed}, or its equivalent form \eqref{LagPertABJM}, is then immediately obtained by generalizing the simpler example of a single chiral multiplet charged under a $U(1)$ gauge field that we presented above.

\subsection{Some group theory}
\label{GROUPTHEORY}

At Chern-Simons level $k=1$, the $SU(4)_R \times U(1)_b$ internal symmetry of ABJM theory is enhanced to $SO(8)_R$, so we construct the gravity dual as a consistent truncation of $\cn=8$ supergravity.  It is important to pair up the operators ${\cal O}_B^\alpha$ and ${\cal O}_F^\alpha$ with a subset of the 70 scalars and pseudo-scalars of the supergravity theory.  The triality of $SO(8)$ complicates the search for the correct correspondence.  A group theory argument to solve this problem is presented in this section.  It is a subtle argument, but it is not necessary to follow it closely in a first reading of this paper.  The reader can pause to consider the result in \reef{Correspondence} below and then proceed to the next section.

Since we want to construct the gravity dual of the perturbation \eqref{LagPertABJM}, we need to know which supergravity fields correspond to the operators ${\cal O}_B^\alpha$ and ${\cal O}_F^\alpha$.  The $35$ scalars and $35$ pseudo-scalars of ${\cal N} = 8$ gauged supergravity transform, respectively, in the ${\bf 35}_v$ and ${\bf 35}_c$ representations of $SO(8)_R$.  Correspondingly, in the undeformed ABJM theory, the operators ${\cal O}_B^\alpha$ and ${\cal O}_F^\alpha$ are states in the ${\bf 35}_v$ and ${\bf 35}_c$ representations of the $SO(8)_R$ symmetry, and according to the AdS/CFT dictionary each of them must be dual to a bulk scalar or pseudo-scalar field.  We match the bulk fields with the boundary operators by considering their transformation properties under a subgroup of $SO(8)_R$ that is broken by the perturbation \eqref{LagPertABJM}.  The simplest such subgroup is $U(1)_R \times SU(2) \times SU(2)\times U(1)_b$, which we encountered before as the symmetry group acting on the ${\cal N} = 2$ super-fields.

We identify $U(1)_R \times SU(2) \times SU(2)\times U(1)_b$ as a subgroup of $SO(8)_R$ by first considering $SU(4)_R \times U(1)_b$ as the subgroup of $SO(8)_R$
under which
 \es{SO8Decomp}{
  {\bf 8}_v &\to {\bf 4}_1 \oplus {\bf \bar 4}_{-1} \,, \\
  {\bf 8}_c &\to {\bf \bar 4}_1 \oplus {\bf 4}_{-1} \,, \\
  {\bf 8}_s &\to {\bf 6}_0 \oplus {\bf 1}_2 \oplus {\bf 1}_{-2} \,,
 }
and then embedding $U(1)_R \times SU(2) \times SU(2)$ into $SU(4)_R$ such that
  \es{SU4Decomp}{
  {\bf 4} &\to  ({\bf 2}, {\bf 1})_{\frac 12}
   \oplus ({\bf 1}, {\bf 2})_{-\frac 12} \,, \\
  {\bf \bar 4} &\to  ({\bf 2}, {\bf 1})_{-\frac 12} 
   \oplus ({\bf 1}, {\bf 2})_{\frac 12} \,, \\
  {\bf 6} &\to ({\bf 2}, {\bf 2})_{0} \oplus ({\bf 1}, {\bf 1})_1\oplus ({\bf 1}, {\bf 1})_{-1} \,.
 }
(From here on we make no distinction between the ${\bf 2}$ and ${\bf \bar 2}$ of $SU(2)$.)  The justification of \eqref{SO8Decomp}--\eqref{SU4Decomp} is as follows.  As mentioned above, if we write ABJM theory at level $k$ in super-field components the global symmetry $SU(4)_R \times U(1)_b$ becomes manifest.  One finds that $Y^A = (Z^a, W^{\dagger b})$ transforms in the ${\bf 4}_1$ of $SU(4)_R \times U(1)_b$, $\psi_A = (\epsilon_{ac} \chi^c, \epsilon_{bd} \eta^{\dagger d})$ transforms in the ${\bf \bar 4}_{1}$, and the six supercharges corresponding to the ${\cal N} = 6$ manifest supersymmetry transform in the ${\bf 6}_0$.  When $k=1$, $SU(4)_R \times U(1)_b$ is enhanced to $SO(8)_R$, and the four scalars $Y^A$ and their complex conjugates transform in the ${\bf 8}_v$, while the four fermions $\psi_A$ and their conjugates transform in the ${\bf 8}_c$.  The eight supersymmetries transform in the ${\bf 8}_s$.  What we call ${\bf 8}_v$, ${\bf 8}_c$, and ${\bf 8}_s$ is of course a triality choice, and the choice made in \eqref{SO8Decomp} yields a more immediate comparison with supergravity.

Using \eqref{SO8Decomp} and \eqref{SU4Decomp} and thinking of ${\bf 35}_v$ and ${\bf 35}_c$ as symmetric traceless products of ${\bf 8}_v$ and ${\bf 8}_c$, respectively, it is not hard to see that under $SO(8)_R \to SU(4)_R \times U(1)_b \to U(1)_R \times SU(2) \times SU(2) \times U(1)$ we have
 \es{35vDecomp}{
  {\bf 35}_v &\to {\bf 10}_2 \oplus {\bf \bar {10}}_{-2} \oplus {\bf 15}_0 \to ({\bf 3}, {\bf 1})_{0, 0} \oplus  ({\bf 1}, {\bf 3})_{0, 0} 
    \oplus ({\bf 1}, {\bf 1})_{0, 0} \oplus ({\bf 2}, {\bf 2})_{1, 0} \oplus ({\bf 2}, {\bf 2})_{-1, 0} \oplus \dotsc \,, \\
  {\bf 35}_c &\to {\bf 10}_{-2} \oplus {\bf \bar {10}}_{2} \oplus {\bf 15}_0 \to ({\bf 3}, {\bf 1})_{0, 0} \oplus  ({\bf 1}, {\bf 3})_{0, 0} 
    \oplus ({\bf 1}, {\bf 1})_{0, 0} \oplus ({\bf 2}, {\bf 2})_{1, 0} \oplus ({\bf 2}, {\bf 2})_{-1, 0} \oplus \dotsc \,,
}   
where the indices are the $U(1)_R$ and $U(1)_b$ charges, and from the last expression on each line we omitted the terms that have non-zero $U(1)_b$ charge.  See Table~\ref{SomeScalarOperators} for explicit expressions of the scalar and pseudo-scalar operators corresponding to the various terms on the right-hand side of \eqref{35vDecomp}.  
\begin{table}[htdp]
 \begin{center}
 \begin{tabular}{c|c|c}
  $U(1)_R \times SU(2) \times SU(2) \times U(1)_b$ irrep & operator in ${\bf 35}_v$ & operator in ${\bf 35}_c$ \\
  \hline
  \hline
   $({\bf 1}, {\bf 1})_{0, 0}$ & $\tr (Z_c^\dagger Z^c - W^{\dagger c} W_c)$ & $\tr (\chi_c^\dagger \chi^c - \eta^{\dagger c} \eta_c)$  \\
   $({\bf 3}, {\bf 1})_{0, 0}$ & $\tr \left( Z_a^\dagger Z^b - \frac 12 \delta^b_a Z_c^\dagger Z^c \right)$
    & $\tr \left( \chi_a^\dagger \chi^b - \frac 12 \delta^b_a \chi_c^\dagger \chi^c \right)$ \\
   $({\bf 1}, {\bf 3})_{0, 0}$ & $\tr \left( W^{\dagger a} W_b - \frac 12 \delta^a_b W^{\dagger c} W_c \right)$
    & $\tr \left( \eta^{\dagger a} \eta_b - \frac 12 \delta^a_b \eta^{\dagger c} \eta_c \right)$ \\
   $({\bf 2}, {\bf 2})_{1, 0}$ & $\tr \left( Z^a W_b \right)$ & $\tr \left( \chi^\dagger_a \eta^{\dagger b} \right)$ \\
   $({\bf 2}, {\bf 2})_{-1, 0}$ & $\tr \left( Z^\dagger_a W^{\dagger b} \right)$
    & $\tr \left( \chi^a \eta_b \right)$ \\
 \end{tabular}
 \caption{Some explicit formulas for the operators in ${\bf 35}_v$ and ${\bf 35}_c$ corresponding to the decomposition \eqref{35vDecomp}.}
  \label{SomeScalarOperators} 
 \end{center}
\end{table}
Using this table, we can now characterize ${\cal O}_B^\alpha$ and ${\cal O}_F^\alpha$.  ~ ${\cal O}_B^1$ (${\cal O}_F^1$) is the state in ${\bf 35}_v$ (${\bf 35}_c$) that corresponds to the singlet $({\bf 1}, {\bf 1})_{0, 0}$ in the decomposition of ${\bf 35}_v$ (${\bf 35}_c$) under $U(1)_R \times SU(2) \times SU(2) \times U(1)_b$.  Under the same decomposition, ${\cal O}_B^2$ (${\cal O}_F^2$) and ${\cal O}_B^3$ (${\cal O}_F^3$) are  states that belong, respectively, to the symmetric and anti-symmetric combinations of $({\bf 1}, {\bf 3})_{0, 0}$ and $({\bf 3}, {\bf 1})_{0, 0}$.  Specifically
 these operators have vanishing charges under the Cartan subalgebra of the product group 
 precisely because they break $SO(8)_R$ to its Cartan subgroup and not any further. 
 
 At the level of linearized perturbations, the ${\cal N} = 8$ supergravity theory contains 35 scalars and 35 pseudo-scalars packaged into the totally antisymmetric tensor $\Sigma_{ijkl}$ that satisfies the duality constraint $(\Sigma^*)^{ijkl} = \epsilon^{ijklmnpq} \Sigma_{mnpq}$.  The indices of $\Sigma_{ijkl}$ are ${\bf 8}_s$ indices.  The real and imaginary parts of $\Sigma_{ijkl}$ have different duality properties and transform in different $35$-dimensional representations of $SO(8)$.  We can take the self-dual real part to transform in the ${\bf 35}_v$, and the anti-self-dual imaginary part to transform in the ${\bf 35}_c$.

We can take the $U(1)_R$ subgroup of $SO(8)_R$ to be given by $SO(2)$ rotations of the $12$ indices, and $U(1)_b$ to correspond to $SO(2)$ rotations of the $34$ indices.  This assignment is consistent with the decomposition of the eight supercharges in the ${\bf 8}_s$ under $U(1)_R \times SU(2) \times SU(2) \times U(1)_b$---see \eqref{SO8Decomp} and \eqref{SU4Decomp}.  The $SU(2) \times SU(2) \cong SO(4)$ factor acts by rotating the remaining $5678$ indices.  

The operators ${\cal O}_B^\alpha$ and ${\cal O}_F^\alpha$ have vanishing $U(1)_R$ and $U(1)_b$ charges, so we should examine which of the $\Sigma_{ijkl}$ have the same property.  These are the complex field $\Sigma_{1234}$ and six complex fields $\Sigma_{12ab}$ where $a, b \in \{5, 6, 7, 8\}$ (as well as $\Sigma_{5678}$ and $\Sigma_{34ab}$ with are related to $\Sigma_{1234}$ and $\Sigma_{12ab}$ by duality.)  Clearly, under $U(1)_R \times SU(2) \times SU(2) \times U(1)_b$, $\Sigma_{1234}$ transforms as $({\bf 1}, {\bf 1})_{0, 0}$, and $\Sigma_{12ab}$ transforms as an $SO(4)$ adjoint, which is $ ({\bf 3}, {\bf 1})_{0, 0} \oplus  ({\bf 1}, {\bf 3})_{0, 0}$.

We are now ready to find which $\Sigma_{ijkl}$ correspond to the operators ${\cal O}_B^\alpha$ and ${\cal O}_F^\alpha$.  The result
 \es{Correspondence}{
   {\cal O}_B^1 + i {\cal O}_F^1 \quad &\longleftrightarrow \quad \Sigma_{1234}\,, \qquad\qquad  {\cal O}_B^1 - i {\cal O}_F^1 \quad \longleftrightarrow \quad \Sigma_{5678} \,, \\
   {\cal O}_B^2 + i {\cal O}_F^2 \quad &\longleftrightarrow \quad \Sigma_{1256}\,, \qquad\qquad  {\cal O}_B^2 - i {\cal O}_F^2 \quad \longleftrightarrow \quad \Sigma_{3478} \,, \\
   {\cal O}_B^3 + i {\cal O}_F^3 \quad &\longleftrightarrow \quad \Sigma_{1278}\,, \qquad\qquad  {\cal O}_B^3 - i {\cal O}_F^3 \quad \longleftrightarrow \quad \Sigma_{3456} \,, 
 }
is shown as follows.  As explained above, ${\cal O}_B^1$ and ${\cal O}_F^1$ are singlets under $U(1)_R \times SU(2) \times SU(2) \times U(1)_b$.  They should correspond to $\Sigma_{1234}$.  ${\cal O}_B^2$ and ${\cal O}_F^2$ (${\cal O}_B^3$ and ${\cal O}_F^3$) correspond to the symmetric (anti-symmetric) combination of $ ({\bf 3}, {\bf 1})$ and $({\bf 1}, {\bf 3})$.  In the fundamental representation of $SO(4)$ the Cartan generators can be taken to be $\diag \{1, -1, 1, -1 \}$ (as the zero weight state in $({\bf 3}, {\bf 1})$) and $\diag \{1, -1, -1, 1 \}$ (as the zero weight state in $({\bf 1}, {\bf 3})$).  This implies that the symmetric (anti-symmetric) combination of $ ({\bf 3}, {\bf 1})$ and $({\bf 1}, {\bf 3})$ is therefore given by $\Sigma_{1256}$ ($\Sigma_{1278}$).  Taking into account the duality property of $\Sigma_{ijkl}$, the correspondence in \eqref{Correspondence} follows.

\section{Consistent truncation of ${\cal N} = 8$ gauged supergravity}
\label{TRUNCATION}

\subsection{An ${\cal N}=2$ truncation}

The field content of the ${\cal N} = 8$ theory consists of a four-dimensional metric $g_{\mu\nu}$, 8 gravitinos $\psi^i_\mu$ (where $i$ is an $SO(8)$ spinor index in ${\bf 8}_s$), $28$ gauge fields $A_\mu^{ij}$ (antisymmetric in the ${\bf 8}_s$ indices $ij$), 56 Majorana dilatinos $\chi^{ijk}$ (antisymmetric in the ${\bf 8}_s$ indices $ijk$), and 35 scalars and 35 pseudo-scalars packaged into a 56-bein ${\cal V}$.  We use the gauge where the $56$-bein is written as
 \es{56bein}{
  {\cal V} = \exp \begin{pmatrix}
    0 & \Sigma_{ijkl} \\
    (\Sigma^*)^{ijkl} & 0  
  \end{pmatrix} \,,
 }
with $\Sigma_{ijkl}$  a totally antisymmetric complex field satisfying the self duality condition  $(\Sigma^*)^{ijkl} = \epsilon^{ijklmnpq} \Sigma_{mnpq}$ that we introduced in section~\ref{GROUPTHEORY}.  In this section we work in Lorentzian signature.  

Since the perturbed Lagrangian \eqref{LagABJMDeformed} preserves a $U(1)_R \times U(1)^3$ subgroup of $SO(8)$, so should its supergravity dual.  We can find the supergravity dual by first restricting the ${\cal N} = 8$ theory to its $U(1)^3$-invariant sector,\footnote{We do not require invariance of the truncated theory under $U(1)_R$.  If we did that, we would eliminate the supersymmetric parters of the fields with zero R-charge, and we would not have a supergravity theory.} and then writing down and solving the BPS equations.  As we explained in section~\ref{GROUPTHEORY}, we take $U(1)^3$ to act as $SO(2)$ rotations in the $34$, $56$, and $78$ indices.  The fields that survive the truncation are the metric $g_{\mu\nu}$, 2 gravitinos $\psi^1_\mu$ and $\psi^2_\mu$, 4 gauge fields $A_\mu^{12}$, $A_\mu^{34}$, $A_\mu^{56}$, $A_\mu^{78}$, and 3 complex scalars $\Sigma_{1234}$, $\Sigma_{1256}$, and $\Sigma_{1278}$.  All the other fields of the ${\cal N} = 8$ gauged supergravity theory are set to zero.  The truncated theory is an ${\cal N} = 2$ theory with one graviton multiplet and 3 vector multiplets.

We now work out the bosonic Lagrangian of this ${\cal N} = 2$ truncation.  The four  gauge fields are dropped because they play no role in the classical solutions we seek.\footnote{It is consistent to set the gauge fields to zero because there are no charged matter fields in our ${\cal N} = 2$ truncation.}  We take
 \es{SigmaTruncate}{
  \Sigma_{1234} = \rho_1 e^{i \theta_1} \,, \qquad
   \Sigma_{5678} = \rho_1 e^{-i\theta_1} \,, \\
  \Sigma_{1256} = \rho_2 e^{i \theta_2} \,, \qquad
   \Sigma_{3478} = \rho_2 e^{-i\theta_2} \,, \\
  \Sigma_{1278} = \rho_3 e^{i \theta_3} \,, \qquad
   \Sigma_{3456} = \rho_3 e^{-i\theta_3} \,, \\  
 }
and we set all the other independent components of $\Sigma_{ijkl}$ to zero.  It is convenient to define the three complex fields
 \es{zDef}{
  z^\alpha = e^{i \theta_\alpha} \tanh \rho_\alpha  \,.
 }

The bosonic part of the action, with the $SO(8)$ tensors taken from \cite{deWit:1982ig}, is
 \es{Sbosonic}{
  S_b =\frac{1}{8 \pi G_4} \int d^4 x\, \sqrt{-g} \left[ \frac 12 R - \frac{1}{96} {\cal A}_\mu^{ijkl} {\cal A}^\mu_{ijkl} + g^2 \left(\frac 34 \abs{A_1^{ij}}^2 - \frac{1}{24} \abs{A_{2jkl}^i}^2 \right) \right]\,,
 }
where $G_4$ is the Newton constant in four dimensions.   With the ansatz \eqref{SigmaTruncate} a tedious computation yields 
 \es{SbosonicFinal}{
  S_b   &=  \frac{1}{8 \pi G_4} \int d^4x\, \sqrt{-g} \left[  \frac 12 R - \sum_{\alpha = 1}^3 \frac{\abs{\partial_\mu z^\alpha}^2 }{\left(1 - \abs{z^\alpha}^2 \right)^2} 
   + \frac{1}{L^2} \left(- 3 + \sum_{\alpha = 1}^3 \frac{2}{1 - \abs{z^\alpha}^2} \right) \right]\,,
 }
where instead of using the gauge coupling constant $g$ we introduced the length scale $L$, normalized such that an extremum of this action is $AdS_4$ with radius of curvature $L$.   The action for the scalar fields in \eqref{SbosonicFinal} is a sigma model action with $\HH^2 \times \HH^2 \times \HH^2$ target space.

This action is not new.  The kinetic term for the scalars is familiar from the $stu$-model where it is determined by an ${\cal N} = 2$ prepotential.  The scalar potential is then fixed by ${\cal N} = 2$ supersymmetry in the sense that it can be derived from the same prepotential:  see, for example, section~3.2 of \cite{Cacciatori:2009iz}.

\subsection{An ${\cal N}=1$ formulation}

Since the gauge fields in the ${\cal N}=2$ truncation of the previous section are not needed, it is convenient to pass to an ${\cal N}=1$ description.  The ${\cal N} = 2$ theory consists of a graviton multiplet and three vector multiplets, which in ${\cal N} = 1$ language should be written as one graviton, one gravitino, three vector, and three chiral multiplets.  From now on we will ignore the gravitino and vector multiplets, and work effectively in an ${\cal N} = 1$ supergravity theory with a graviton multiplet (consisting of the metric and a gravitino) and three chiral multiplets (consisting of a complex scalar and a Majorana fermion each).
The discussion of  $\cn = 1$ supergravity is based on Chapter 18 of \cite{Freedman:2012zz}.

For an ${\cal N} = 1$ supergravity theory with chiral matter, the bosonic action can be written as
 \es{Lag}{
  S_b = \frac{1}{8 \pi G_4} \int d^4 x\, \sqrt{-g} \left[ \frac{1}{2}R  - {\cal K}_{\alpha \bar{\beta}} \partial_\mu z^\alpha \partial^\mu \bar z^{\bar \beta} - V_F \right] \,,
 }
where ${\cal K}_{\alpha {\bar \beta}}$ is the K\"ahler metric, and $V_F$ is the potential.  The K\"ahler metric can be obtained from a K\"ahler potential ${\cal K}$ from ${\cal K}_{\alpha \bar \beta} = \partial_\alpha \partial_{\bar \beta} {\cal K}$, and the potential $V_F$ can be written in terms of the superpotential $W$ as
 \es{VFGeneral}{
  V_F = e^{\cal K} \left(-3 W \bar W + \nabla_\alpha W {\cal K}^{\alpha \bar \beta} \nabla_{\bar \beta} \bar W \right)\,,
 }
where the K\"ahler covariant derivative is defined as $\nabla_\alpha W = \partial_\alpha W + (\partial_\alpha {\cal K}) W$.  

For us, the K\"ahler potential and K\"ahler metric are
 \es{Kahler}{
  {\cal K} = - \sum_{\alpha=1}^3 \log \left[ \left( 1 - \abs{z^\alpha}^2 \right) \right]\,, \qquad
   {\cal K}_{\alpha {\bar \beta}} = \frac{\delta_{\alpha {\bar \beta}}}{(1 - \abs{z^\alpha}^2)^2} \,,
 } 
and the superpotential is 
 \es{Superpot}{
  W = \frac{1 + z^1 z^2 z^3}{L} \,.
 }
Then using \eqref{VFGeneral} with 
 \es{nablaW}{
  \nabla_\alpha W = \frac{\bar z^\alpha + z^1 z^2 z^3 / z^\alpha}{1 - \abs{z^\alpha}^2 } \,, \qquad
    \nabla_{\bar \alpha} \bar W = \frac{z^{\bar \alpha} + \bar z^1 \bar z^2 \bar z^3 / \bar z^{\bar \alpha}}{1 - \abs{ z^{\bar \alpha}}^2} \,,
 }
we obtain
 \es{GotVF}{
  V_F = \frac{1}{L^2} \left(3 - \sum_{\alpha = 1}^3 \frac{2}{1 - \abs{z^\alpha}^2} \right) \,.
 }
The normalization of the potential was chosen so that the $AdS_4$ extremum of \eqref{Lag} obtained when $z^\alpha  = \bar z^\alpha \equiv 0$ has radius $L$.
It is striking that the coupled cubic $W$ produces a decoupled $V_F$.

In Lorentzian signature the fermionic partners of the metric and the complex fields $z^\alpha$ are a Majorana gravitino $\hat \psi_\mu$ and three Majorana fermions $\hat \chi^\alpha$, the hats signifying four-component fermions.  The linearized supersymmetry variations, with supersymmetry parameter $\hat \epsilon$ also satisfying the Majorana condition, are
 \es{SUSYvariations}{
  \delta P_L \hat \psi_\mu &= \left( \partial_\mu + \frac{1}{4} \omega_\mu{}^{ab} \gamma_{ab} - \frac{3}2 i {\cal A}_\mu \right) P_L \hat \epsilon
   + \frac 12 \gamma_\mu e^{{\cal K}/2} W P_R \hat \epsilon  \,, \\
    \delta P_R \hat \psi_\mu &= \left( \partial_\mu + \frac{1}{4} \omega_\mu{}^{ab} \gamma_{ab} + \frac{3}2 i {\cal A}_\mu \right) P_R \hat \epsilon
   + \frac 12 \gamma_\mu e^{{\cal K}/2} \bar{W} P_L \hat \epsilon  \,, \\ 
   \delta P_L \hat \chi^\alpha &=  P_L \left(\slashed{\partial} z^\alpha - e^{{\cal K}/2}  g^{\alpha \bar\beta} \nabla_{\bar \beta} \bar W \right) 
   \hat \epsilon \,, \\
    \delta P_R \hat \chi^{\bar \beta} &= P_R \left(\slashed{\partial} \bar z^{\bar \beta} - e^{{\cal K}/2}  g^{\alpha \bar\beta} \nabla_{\alpha} W \right) 
   \hat \epsilon \,,
 }
(see Appendix~\ref{CONVENTIONS} for our conventions) where the K\"ahler connection ${\cal A}_\mu$ is given by
 \es{GotAmu}{
  {\cal A}_\mu =\frac 16 i \left( \partial_\mu z^\alpha {\cal K}_\alpha - 
    \partial_\mu  \bar z^{\bar \alpha} {\cal K}_{\bar \alpha} \right) =   \frac{i}{6} 
    \sum_{\alpha=1}^3 \frac{\bar z^{\bar \alpha} \partial_\mu z^\alpha - z^\alpha \partial_\mu \bar z^{\bar \alpha}}{1 - \abs{z^\alpha}^2} \,.
 }

We can convert the fermions to two-component Weyl spinor notation by writing
 \es{FourToTwo}{
  \hat \psi_\mu = \begin{pmatrix}
   \psi_\mu \\
   \tilde \psi_\mu 
  \end{pmatrix} \,, \qquad
   \hat \chi^\alpha =  \begin{pmatrix}
   \chi^\alpha \\
   \tilde \chi^\alpha
  \end{pmatrix} \,, \qquad
   \hat \epsilon =  \begin{pmatrix}
   \epsilon \\
   \tilde \epsilon
  \end{pmatrix} \,.
 }
The Majorana condition relates the bottom spinor to the complex conjugate of the top one:
 \es{MajoranaCondition}{
  \tilde \psi_\mu = i \sigma_2 (\psi_\mu)^* \,, \qquad
   \tilde \chi^\alpha = i \sigma_2 (\chi^\alpha)^* \,, \qquad
    \tilde \epsilon = i \sigma_2 \epsilon^* \,.
 }
In terms of Weyl spinors, the supersymmetry variations \eqref{SUSYvariations} can be written as
 \es{SUSYvariationsAgain}{
  \delta \psi_\mu &= \left( \partial_\mu + \frac{1}{4} \omega_\mu{}^{ab} \sigma_{[a} \bar \sigma_{b]}  - \frac{3}2 i {\cal A}_\mu \right)  \epsilon
   + \frac 12 \sigma_\mu e^{{\cal K}/2} W \tilde \epsilon  \,, \\
    \delta \tilde \psi_\mu &= \left( \partial_\mu + \frac{1}{4} \omega_\mu{}^{ab} \bar \sigma_{[a} \sigma_{b]} + \frac{3}2 i {\cal A}_\mu \right) \tilde \epsilon
   + \frac 12 \bar \sigma_\mu e^{{\cal K}/2} \bar W \epsilon  \,, \\ 
   \delta \chi^\alpha &= \sigma^\mu \partial_\mu z^\alpha \tilde \epsilon 
     -  e^{{\cal K}/2}  g^{\alpha \bar\beta} \nabla_{\bar \beta} \bar W  \epsilon \,, \\
   \delta \tilde \chi^{\bar \beta}  &= \bar \sigma^\mu \partial_\mu \bar z^{\bar \beta} \epsilon 
     - e^{{\cal K}/2}  g^{\alpha \bar\beta} \nabla_{\alpha} W  \tilde \epsilon \,.
 }
One can show  using the Majorana condition \eqref{MajoranaCondition} that the second and fourth equations are the complex conjugates of the first and third, respectively.

\section{Analytic continuation to Euclidean signature}
\label{CONTINUATION}

As explained in detail in Appendix~\ref{CONVENTIONS}, in Euclidean signature we should not impose the Majorana condition \eqref{MajoranaCondition}.  Instead, the spinors $\tilde \psi_\mu$, $\tilde \chi^\alpha$, and $\tilde \epsilon$ are treated as independent of $\psi_\mu$, $\chi^\alpha$, and $\epsilon$, respectively.  Similarly, we should not require that $\bar z^\alpha$ be the complex conjugate of $z^\alpha$, and we should allow in principle the metric to be complex.  Instead of writing $\bar z^\alpha$, in Euclidean signature we will write $\tilde z^\alpha$.

The bosonic part of the ${\cal N}=1$ Euclidean action is then 
 \es{ActionEuclidean}{
  S_\text{bulk} &=  \frac{1}{8 \pi G_4} \int d^4x\, \sqrt{g} \left[ - \frac 12 R + \sum_{\alpha = 1}^3 \frac{\partial_\mu z^\alpha \partial^\mu \tilde z^\alpha }
    {\left(1 - z^\alpha \tilde z^\alpha \right)^2} 
   +\frac{1}{L^2} \left( 3 - \sum_{\alpha = 1}^3 \frac{2}{1 - z^\alpha \tilde z^\alpha} \right)  \right]\,.
 }
The supersymmetry transformations are the same as \eqref{SUSYvariationsAgain}, with the only exception that we should now use the $\sigma_\mu$ and $\bar \sigma_\mu$ matrices in \eqref{eucsigma}, as appropriate for Euclidean signature.  Plugging in the explicit form of the K\"ahler potential and superpotential from \eqref{Kahler} and \eqref{Superpot}, we obtain
  \es{SUSYvariationsEuclidean}{
  \delta \psi_\mu &= \left( \partial_\mu + \frac{1}{4} \omega_\mu{}^{ab} \sigma_{[a} \bar \sigma_{b]}  + \frac 14  
    \sum_{\alpha=1}^3 \frac{\tilde z^\alpha \partial_\mu z^\alpha - z^\alpha \partial_\mu \tilde z^\alpha}{1 - z^\alpha \tilde z^\alpha} \right)  \epsilon
   + \frac{1 + z^1 z^2 z^3}{2 L \prod_{\beta=1}^3 \sqrt{1 - z^\beta \tilde z^\beta} }  \sigma_\mu \tilde \epsilon  \,, \\
    \delta \tilde \psi_\mu &= \left( \partial_\mu + \frac{1}{4} \omega_\mu{}^{ab} \bar \sigma_{[a} \sigma_{b]} - \frac 14
    \sum_{\alpha=1}^3 \frac{\tilde z^\alpha \partial_\mu z^\alpha - z^\alpha \partial_\mu \tilde z^\alpha}{1 - z^\alpha \tilde z^\alpha} \right) \tilde \epsilon
   + \frac{1 + \tilde z^1 \tilde z^2 \tilde z^3}{2 L \prod_{\beta=1}^3 \sqrt{1 - z^\beta \tilde z^\beta} }  \bar \sigma_\mu \epsilon  \,, \\ 
   \delta \chi^\alpha &=  \sigma^\mu \partial_\mu z^\alpha \tilde \epsilon 
      -  \frac{\left( {1 - z^\alpha \tilde z^\alpha} \right) \left( z^\alpha + \tilde z^1 \tilde z^2 \tilde z^3 / \tilde z^\alpha \right)}
      { \prod_{\beta=1}^3 \sqrt{1 - z^\beta \tilde z^\beta} } \epsilon \,, \\
   \delta \tilde \chi^\alpha &=  \bar \sigma^\mu \partial_\mu \tilde z^\alpha \epsilon -   
      \frac{\left( {1 - z^\alpha \tilde z^\alpha} \right) \left( \tilde z^\alpha + z^1 z^2 z^3 / z^\alpha\right)}
      { \prod_{\beta=1}^3 \sqrt{1 - z^\beta \tilde z^\beta} }   \tilde \epsilon \,.
 }
In the next section we will find BPS solutions that satisfy $\delta \psi_\mu = \delta \tilde \psi_\mu = \delta \chi^\alpha = \delta \tilde \chi^\alpha = 0$.

\section{BPS equations and Killing spinors}
\label{BPS}

\subsection{Metric ansatz and second order equations of motion}

Consider the metric ansatz
 \es{MetricAnsatz}{
  ds^2 = L^2 e^{2 A(r)} ds_{S^3}^2 + e^{2 B(r)} dr^2 
 }
and the vielbein
 \es{Viel}{
  e^i = L e^{A(r)} \hat e^i \,, \qquad
   e^4 = e^{B(r)} dr \,,
 }
where $\hat e^i$ ($i = 1, 2, 3$) is a frame on the unit $S^3$.\footnote{In all of our supergravity calculations we consider an $S^3$ of unit radius.  Since we keep the bulk curvature scale $L$ explicit, the $S^3$ radius can be restored by dimensional analysis when reading off field theory quantities from the supergravity solutions.}  Having two functions $A(r)$ and $B(r)$ in the metric ansatz \eqref{MetricAnsatz} is certainly redundant, but it might be helpful not to specify a gauge just yet.  Later on, we will find it convenient to work in a gauge where the metric is conformally flat:
 \es{Gauge}{
  e^{A(r)} = \frac{r}{L} e^{B(r)} \,, \qquad
   ds^2 = e^{2 B(r)} \left( dr^2 + r^2 d\Omega_3^2 \right) \,.
 }

Let $\hat \omega_{ij}$ be the spin connection on $S^3$.  The spin connection for the frame \eqref{Viel} is
 \es{SpinConnection}{
  \omega^{ij} = \hat \omega^{ij} \,, \qquad
   \omega^{i4} = -\omega^{4i} = A' e^{-B} e^i \,.
 }

The second order equations of motion that follow from extremizing the Euclidean action \eqref{ActionEuclidean} are
 \es{SecondOrder}{
  A'' - A'B' + \sum_{\alpha = 1}^3 \frac{z^\alpha{}' \tilde z^\alpha{}'}{(1 - z^\alpha \tilde z^\alpha)^2} + \frac{e^{-2A + 2B}}{L^2} &= 0 \,, \\
  z^\alpha{}'' + (3 A' - B') z^\alpha{}' + \frac{2 \tilde z^\alpha (z^\alpha{}')^2}{1 - z^\alpha \tilde z^\alpha} + \frac{2z^\alpha e^{2 B}}{L^2} &= 0 \,, \\
  \tilde z^\alpha{}'' + (3 A' - B') \tilde z^\alpha{}' + \frac{2 z^\alpha (\tilde z^\alpha{}')^2}{1 - z^\alpha \tilde z^\alpha} + \frac{2\tilde z^\alpha e^{2 B}}{L^2} &= 0 \,. \\
 }
The Hamiltonian constraint is 
 \es{Hamilt}{
  3(A')^2 - \sum_{\alpha = 1}^3 \frac{z^\alpha{}' \tilde z^\alpha{}'}{(1 - z^\alpha \tilde z^\alpha)^2} - \frac{3 e^{-2A + 2B}}{L^2} + \frac{e^{2B}}{L^2} \left(3 - \sum_{\alpha = 1}^3 \frac{2}{1 - z^\alpha \tilde z^\alpha} \right) = 0 \,.
 }
It can be checked that the Hamiltonian constraint is consistent with the second order equations \eqref{SecondOrder}, meaning that if it holds for a particular value of $r$ the second order equations guarantee that it holds for all $r$.

\subsection{Warm-up:  Supersymmetry of $\HH^4$}

As a warm-up, let's start by solving the BPS equations in the case where $z^\alpha = \tilde z^\alpha = 0$.  We should find that the metric describes $\HH^4$, and we will find the Killing spinors.  The SUSY variations \eqref{SUSYvariationsEuclidean} become
 \es{KillingAdS}{
      \delta \psi_\mu &= \left( \partial_\mu + \frac{1}{4} \omega_\mu{}^{ab} \sigma_{[a} \bar \sigma_{b]} \right) \epsilon
   + \frac 1{2L} \sigma_\mu \tilde \epsilon  \,, \\
       \delta \tilde \psi_\mu &=  \left( \partial_\mu + \frac{1}{4} \omega_\mu{}^{ab} \bar \sigma_{[a} \sigma_{b]}  \right)  \tilde \epsilon
   + \frac 1{2L} \bar \sigma_\mu  \epsilon  \,. \\
 }
Requiring that the two variations in \eqref{KillingAdS} vanish, and specializing for $\mu = i$ and $\mu = r$ we obtain
 \es{KillingExplicit}{
   \nabla_i \begin{pmatrix}
   \epsilon \\
   \tilde \epsilon 
   \end{pmatrix} 
    &=  -i \frac{\sigma_i}2 
     \begin{pmatrix}
      L A' e^{A - B} & -i e^A \\
      -i e^A & -L A' e^{A-B}
     \end{pmatrix}
      \begin{pmatrix}
       \epsilon \\
       \tilde \epsilon 
      \end{pmatrix} \,, \\
  \partial_r \begin{pmatrix}
   \epsilon \\
   \tilde \epsilon 
   \end{pmatrix} 
    &= \frac{i e^B}{2L} \begin{pmatrix}
     0 & 1 \\
     -1 & 0
    \end{pmatrix} 
   \begin{pmatrix}
   \epsilon \\
   \tilde \epsilon 
   \end{pmatrix} \,,
 }
where the derivative $\nabla_i$ appearing in the first equation is the covariant derivative on the unit radius $S^3$.

We expect $\epsilon$ and $\tilde \epsilon$ to be linear combinations of Killing spinors on $S^3$, with coefficients depending on $r$.  These Killing spinors satisfy 
 \es{zetaDef}{
  \nabla_i \zeta = \frac i2 \sigma_i \zeta
 }
(these are invariant under the $SU(2)_\ell$ subgroup of the $SO(4) = SU(2)_\ell \times SU(2)_r$ isometry of $S^3$) and 
 \es{xiDef}{
  \nabla_i \xi = - \frac i2 \sigma_i \xi
 }
(these are invariant under $SU(2)_r$).  So the eigenvalues of the matrix appearing in the first line of \eqref{KillingExplicit} should be $\pm 1$, or in other words the determinant of that matrix should be $-1$:
 \es{detMatrix}{
  -L^2 (A')^2 e^{2 (A- B)} + e^{2A} = -1 \,.
 }
In the conformally flat gauge \eqref{Gauge}, this equation becomes $r (e^{-A})' =  \sqrt{ 1 + e^{-2A}}$, whose solution is $e^{-A} = \frac{r_0}{2 r} - \frac{r}{2 r_0}$, where $r_0$ is an integration constant that can be set to $r_0 = 1$ by reparameterizing $r$.   The metric can therefore be written as
 \es{MetricConformal}{
  ds^2 = \frac{4 L^2}{\left(1 - r^2 \right)^2} \left( dr^2 +  r^2 d\Omega_3^2 \right) \,.
 }
This is the metric on $\HH^4$ of curvature radius $L$, where the unit $\HH^4$ is parameterized as the Poincar\'e disk of unit radius.

We still need to check that all the equations in \eqref{KillingExplicit} can be solved consistently.  In our gauge, \eqref{KillingExplicit} becomes
  \es{KillingExplicitGauge}{
   \nabla_i \begin{pmatrix}
   \epsilon \\
   \tilde \epsilon 
   \end{pmatrix} 
    &=  -i \frac{\sigma_i}{2} 
     \begin{pmatrix}
       \frac{1+r^2}{1 - r^2} & - \frac{2i r}{1 - r^2}  \\
      - \frac{2ir}{1 - r^2}   & - \frac{1+r^2}{1 - r^2} 
     \end{pmatrix}
      \begin{pmatrix}
       \epsilon \\
       \tilde \epsilon 
      \end{pmatrix} \,, \\
  \partial_r \begin{pmatrix}
   \epsilon \\
   \tilde \epsilon 
   \end{pmatrix} 
    &= \frac{i}{1-r^2} \begin{pmatrix}
     0 & 1 \\
     -1 & 0
    \end{pmatrix} 
   \begin{pmatrix}
   \epsilon \\
   \tilde \epsilon 
   \end{pmatrix} \,.
 }
It can be checked that these equations are solved by
 \es{KillingH4Soln1}{
  \begin{pmatrix} \epsilon \\
  \tilde \epsilon 
  \end{pmatrix} 
   = \frac{1}{\sqrt{1-r^2}} \begin{pmatrix}
      r \\
      -i 
   \end{pmatrix}
    \zeta 
 } 
 or  
 \es{KillingH4Soln2}{
  \begin{pmatrix} \epsilon \\
  \tilde \epsilon 
  \end{pmatrix} 
   = \frac{1}{\sqrt{1-r^2}} \begin{pmatrix}
       1\\
       -i r
   \end{pmatrix} 
    \xi \,.
 }
This is a 4-dimensional space of solutions because on $S^3$ there exist two linearly independent solutions to $\nabla_i \zeta = \frac i2 \sigma_i \zeta$ and two linearly independent solutions to $\nabla_i \xi = - \frac i2 \sigma_i \xi$.  The Killing spinors in \eqref{KillingH4Soln1}--\eqref{KillingH4Soln2} are parameters for the odd generators of the Euclidean ${\cal N} =1$ superconfromal algebra $OSp(1|2, 2)$, which is the Euclidean continuation of the Lorentzian superconformal algebra $OSp(1|4)$.   The even subalgebra of $OSp(1|2, 2)$ is the conformal algebra $USp(2, 2) \cong SO(4, 1)$.

When the ${\cal N} = 1$ supergravity theory whose $\HH^4$ solution we just derived is embedded in the ${\cal N} = 8$ gauged supergravity theory, there are $8$ independent supersymmetry transformations with parameters $(\epsilon^i, \tilde \epsilon^i)$, $i = 1, \ldots, 8$ that transform in the spinor ${\bf 8}_s$ representation of the $SO(8)_R$ symmetry.  Each of these parameters can be chosen to be \eqref{KillingH4Soln1} or \eqref{KillingH4Soln2}, yielding the 32 odd generators of $OSp(8 | 2, 2)$.

 \subsection{BPS equations for the three-scalar system}
 
We return to the full system of BPS conditions obtained from 
 the vanishing of the supersymmetry variations \eqref{SUSYvariationsEuclidean}. With our metric and frame \eqref{MetricAnsatz}--\eqref{Viel},  these conditions can be written as the system of equations:
  \es{KillingRotInv1}{
   \nabla_i \begin{pmatrix}
   \epsilon \\
   \tilde \epsilon 
   \end{pmatrix} 
    &=  -i \frac{\sigma_i}2 
     \begin{pmatrix}
      L A' e^{A - B} & -i \frac{e^A (1 + z_1 z_2 z_3)}{\sqrt{\left(1 - z_1 \tilde z_1 \right) \left(1 - z_2 \tilde z_2 \right) \left(1 - z_3 \tilde z_3 \right)}} \\
      -i \frac{e^A(1 + \tilde z_1 \tilde z_2 \tilde z_3)}{\sqrt{\left(1 - z_1 \tilde z_1 \right) \left(1 - z_2 \tilde z_2 \right) \left(1 - z_3 \tilde z_3 \right)}} & -L A' e^{A-B}
     \end{pmatrix}
      \begin{pmatrix}
       \epsilon \\
       \tilde \epsilon 
      \end{pmatrix} 
     \,, 
   }
  \es{KillingRotInv2}{  
  \partial_r \begin{pmatrix}
   \epsilon \\
   \tilde \epsilon 
   \end{pmatrix} 
    &= \Biggl[ \frac{i e^B}{2L \prod_{\alpha=1}^3 \sqrt{1 - z_\alpha \tilde z_\alpha}} \begin{pmatrix}
     0 & 1 + z_1 z_2 z_3 \\
     -(1 + \tilde z_1 \tilde z_2 \tilde z_3)  & 0
    \end{pmatrix} \\
    &\qquad\qquad\qquad\qquad\qquad\qquad\qquad- \frac{1}4\sum_{\alpha=1}^3 \frac{\tilde z_\alpha \partial_r z_\alpha - z_\alpha \partial_r \tilde z_\alpha}{1 - z_\alpha \tilde z_\alpha} 
     \begin{pmatrix} 1 & 0 \\ 0 & -1 \end{pmatrix} \Biggr]
   \begin{pmatrix}
   \epsilon \\
   \tilde \epsilon 
   \end{pmatrix}
     \,, 
   }
   \es{KillingRotInv3}{  
   \begin{pmatrix} 0 \\ 0 \end{pmatrix} &=
    \left[ - \frac 1L \frac{\left( {1 - z_\alpha \tilde z_\alpha} \right) }
      {\prod_{\beta=1}^3 \sqrt{1 - z_\beta \tilde z_\beta} }  \begin{pmatrix}
    z_\alpha + \frac{\tilde z_1 \tilde z_2 \tilde z_3}{\tilde z_\alpha} & 0 \\
    0 & \tilde z_\alpha + \frac{z_1 z_2 z_3}{z_\alpha}
   \end{pmatrix} 
    + e^{-B} \begin{pmatrix} 0 & -i \partial_r z_\alpha \\ i \partial_r \tilde z_\alpha & 0  \end{pmatrix}
        \right]
   \begin{pmatrix}
   \epsilon \\
   \tilde \epsilon 
   \end{pmatrix} \,.
 }

We expect to find non-trivial solutions where $\epsilon$ and $\tilde \epsilon$ are proportional either to the $S^3$ left-invariant Killing spinors $\zeta$ or to the right-invariant ones $\xi$.  The first equation above, \eqref{KillingRotInv1}, then simplifies to
 \es{zetaProp}{
  \begin{pmatrix}
   0 \\ 0  
   \end{pmatrix} = \begin{pmatrix}
      L A' e^{A - B} \pm 1 & -i \frac{e^A (1 + z_1 z_2 z_3)}{\prod_{\beta=1}^3 \sqrt{1 - z^\beta \tilde z^\beta}} \\
      -i \frac{e^A(1 + \tilde z_1 \tilde z_2 \tilde z_3)}{\prod_{\beta=1}^3 \sqrt{1 - z^\beta \tilde z^\beta}} & -L A' e^{A-B} \pm 1
     \end{pmatrix}
      \begin{pmatrix}
       \epsilon \\
       \tilde \epsilon 
      \end{pmatrix} \,,
 }
where the upper signs correspond to $\zeta$ and the lower signs to $\xi$.  For each choice of sign, eqs.~\eqref{zetaProp} and \eqref{KillingRotInv3} are eight linear algebraic equations for $\epsilon$ and $\tilde \epsilon$ that must be consistent with one another.  The consistency conditions can be found by solving for $\tilde \epsilon$ in terms of $\epsilon$ from one of the equations and plugging the solution into the other ones.  Equivalently, one can combine \eqref{KillingRotInv3} with \eqref{zetaProp} into a system of eight equations for two unknowns $\epsilon$ and $\tilde \epsilon$;  the system has nontrivial solutions if and only if all the $2\times 2$ minors have zero determinant.  From these conditions, we find the BPS equations:
 \es{ConsistencyConditions}{
  L e^{A-B} \left(1 + \tilde z^1 \tilde z^2 \tilde z^3 \right) z^\alpha{}'
    &= \left(\pm 1 - L e^{A-B} A' \right) \left(1 - z^\alpha \tilde z^\alpha \right) \left(z^\alpha + \frac{\tilde z^1 \tilde z^2 \tilde z^3}{\tilde z^\alpha} \right) \,, \\
  L e^{A-B} \left(1 + z^1 z^2 z^3 \right) \tilde z^\alpha{}'
    &= \left(\mp 1 - L e^{A-B} A' \right) \left(1 - z^\alpha \tilde z^\alpha \right) \left(\tilde z^\alpha + \frac{z^1 z^2 z^3}{z^\alpha} \right)  \,, \\
  -1&= -L^2 (A')^2 e^{2A-2B} + e^{2 A} \frac{(1 + z^1 z^2 z^3)(1 + \tilde z^1 \tilde z^2 \tilde z^3)}{\prod_{\beta=1}^3 (1 - z^\beta \tilde z^\beta) } \,.
 }
The first equation in \eqref{ConsistencyConditions} is obtained from the bottom row of \eqref{zetaProp} and the top row  of \eqref{KillingRotInv3}.  The second equation is obtained from the top row of \eqref{zetaProp} and the bottom row  of \eqref{KillingRotInv3}.  Lastly, the third equation in \eqref{ConsistencyConditions} is obtained from \eqref{zetaProp}.  It can be checked that \eqref{ConsistencyConditions} imply the second order equations of motion \eqref{SecondOrder} and the Hamiltonian constraint \eqref{Hamilt}, as well as the vanishing of the other $2 \times 2$ minors of the system of equations for $\epsilon$ and $\tilde \epsilon$.

We can analyze the equations \eqref{ConsistencyConditions} in the conformally flat gauge \eqref{Gauge} where $L e^{A-B} = r$.   A regular metric at $r=0$ would have $e^A = r + O(r^2)$, which implies that $1 - L e^{A-B} A' $ vanishes at $r=0$, but $-1 - L e^{A-B} A'$ does not.  Since the left-hand sides of the first two equations in \eqref{ConsistencyConditions} also vanish at $r=0$ if one assumes that $z^\alpha$ and $\tilde z^\alpha$ are regular, these two equations imply that $z^\alpha(0) \tilde z^\alpha(0) = -z^1(0) z^2(0) z^3(0)$ for the upper choice of signs, and $z^\alpha(0) \tilde z^\alpha(0) = -\tilde z^1(0) \tilde z^2(0) \tilde z^3(0)$ for the lower choice of signs.  Developing a series solution to \eqref{ConsistencyConditions} around $r=0$, one can see that $z^\alpha(r)$ and $\tilde z^\alpha(r)$ are all proportional:
 \es{Proportionality}{
  z^\alpha(r) = z^\alpha(0) f(r) \,, \qquad \tilde z^\alpha(r) = \tilde z^\alpha(0) f(r) \,,
 } 
for some function $f$ satisfying $f(0) = 1$.

Let's focus on the upper choice of signs.  Denoting $z^\alpha(0) = c_\alpha$ and using $z^\alpha(0) \tilde z^\alpha(0) = -z^1(0) z^2(0) z^3(0)$, we can write
 \es{UpperSigns}{
  z^\alpha(r) = c_\alpha f(r) \,, \qquad \tilde z^\alpha(r) = - \frac{c_1 c_2 c_3}{c_\alpha} f(r) \,.
 }
Plugging \eqref{UpperSigns} in the first two equations of \eqref{ConsistencyConditions} and eliminating $A'$, one obtains a differential equation for $f$:
 \es{feq}{
  f' = \frac{2 (f - 1) (c_1 c_2 c_3 f + 1)}{(1 + c_1 c_2 c_3) r} \,.
 }
The solution of this first order differential equation is
 \es{fSoln}{
  f(r) = \frac{1 - (r/r_0)^2}{1 + c_1 c_2 c_3 (r/r_0)^2} \,,
 } 
where $r_0$ is an integration constant.  We can set $r_0 = 1$ by reparameterizing $r$, and we will henceforth do so.  Using \eqref{UpperSigns} and \eqref{fSoln}, one can find an algebraic equation for $A$ from substituting the expression for $A'$ found from the first equation in \eqref{ConsistencyConditions} into the last equation of \eqref{ConsistencyConditions}.  Solving this algebraic equation yields the metric
 \es{GotMetric}{
  ds^2 = \frac{4 L^2 (1 + c_1 c_2 c_3) (1 + c_1 c_2 c_3 r^4) }{(1-r^2)^2 (1 + c_1 c_2 c_3 r^2)^2} \left( dr^2 + r^2 d\Omega_3^2 \right) \,.
 }

To summarize, for the upper choice of signs in \eqref{zetaProp}, we obtain a three-parameter family of solutions
 \es{zUpper}{
  \text{$SU(2)_\ell \times OSp(2|2)_r$ branch:} \quad 
    ds^2 &= \frac{4 L^2 (1 + c_1 c_2 c_3) (1 + c_1 c_2 c_3 r^4) }{(1-r^2)^2 (1 + c_1 c_2 c_3 r^2)^2} \left( dr^2 + r^2 d\Omega_3^2 \right) \,, \\
    z^\alpha &= \frac{c_\alpha (1 - r^2)}{1 + c_1 c_2 c_3 r^2} \,, \qquad
      \tilde z^\alpha = - \frac{c_1 c_2 c_3 (1 - r^2)}{c_\alpha ( 1 + c_1 c_2 c_3 r^2) } \,.
 }
(The label ``$SU(2)_\ell \times OSp(2|2)_r$ branch'' will be explained shortly.)  The Killing spinors should of course be proportional to the left-invariant $S^3$ Killing spinors $\zeta$ satisfying $\nabla_i \zeta = \frac{i}{2} \sigma_i \zeta$ because that's the dependence that led us to consider the upper choices of sign.  From any of the equations in \eqref{KillingRotInv3} and \eqref{zetaProp} we can moreover find that $\epsilon = i r \tilde \epsilon$ by simply plugging in the solution \eqref{GotMetric}--\eqref{zUpper}.  The explicit $r$-dependence of the Killing spinors can be found by solving \eqref{KillingRotInv2}, which is an equation that we haven't considered so far.  We find that the solution is
 \es{KillingUpper}{
  \begin{pmatrix}
   \epsilon \\
   \tilde \epsilon
  \end{pmatrix}
   = \frac{\left(1 + c_1 c_2 c_3 r^4 \right)^{1/4}}{\sqrt{(1 - r^2) \left(1 + c_1 c_2 c_3 r^2\right)}}\begin{pmatrix}
    r \\ -i
   \end{pmatrix} \zeta \,.
 }

These Killing spinors are the fermionic parameters of an $OSp(1|2)$ algebra whose bosonic sub-algebra is the  $SU(2)_r$ subgroup of the $SO(4) \cong SU(2)_\ell \times SU(2)_r$ isometry group of $S^3$.  Within the ${\cal N} = 1$ supergravity theory we considered, our solution therefore has $SU(2)_\ell \times OSp(1|2)_r$ symmetry, where the subscript $r$ on $OSp(1|2)_r$ means that this group contains $SU(2)_r$ as opposed to $SU(2)_\ell$.  However, we obtained our ${\cal N} = 1$ theory from an ${\cal N} = 2$ truncation of ${\cal N} = 8$ gauged supergravity, and in the ${\cal N} = 2$ theory we have supersymmetry transformations with two independent parameters $(\epsilon^i, \tilde \epsilon^i)$ with $i = 1, 2$, transforming in the fundamental of an $SO(2)_R$ symmetry group.  Seen as extrema of the ${\cal N} = 2$ gauged supergravity action, our backgrounds \eqref{zUpper} are invariant under $SU(2)_\ell \times OSp(2|2)_r$, which justifies the label we gave to this branch of solutions in \eqref{zUpper}.

One can go through a similar exercise for the lower choice of signs in \eqref{KillingRotInv1} to find solutions that preserve $OSp(2|2)_\ell \times SU(2)_r$.  As can be seen by examining the BPS equations \eqref{ConsistencyConditions}, one simply exchanges $z^\alpha$ with $\tilde z^\alpha$ in this case:
 \es{zLower}{
    \text{$OSp(2 | 2)_\ell \times SU(2)_r$ branch:}\quad   
     ds^2 &= \frac{4 L^2 (1 + c_1 c_2 c_3) (1 + c_1 c_2 c_3 r^4) }{(1-r^2)^2 (1 + c_1 c_2 c_3 r^2)^2} \left( dr^2 + r^2 d\Omega_3^2 \right) \,, \\
    z^\alpha &= - \frac{c_1 c_2 c_3 (1 - r^2)}{c_\alpha ( 1 + c_1 c_2 c_3 r^2) }\,,  \qquad
       \tilde z^\alpha = \frac{c_\alpha (1 - r^2)}{1 + c_1 c_2 c_3 r^2} \,.
 }
It is also straightforward to calculate the Killing spinors
 \es{KillingLower}{
  \begin{pmatrix}
   \epsilon \\
   \tilde \epsilon
  \end{pmatrix}
   = \frac{\left(1 + c_1 c_2 c_3 r^4 \right)^{1/4}}{\sqrt{(1 - r^2) \left(1 + c_1 c_2 c_3 r^2\right)}}\begin{pmatrix}
    1 \\ -ir
   \end{pmatrix} \xi \,,
 }
which are now the fermionic parameters of an $OSp(1|2)_\ell$ algebra containing $SU(2)_\ell$.  In the ${\cal N} =2$ supergravity theory there are again two independent supersymmetry variations, and the background \eqref{zLower} is invariant under $OSp(2|2)_\ell \times SU(2)_r$.

\section{Field theory interpretation}
\label{HOLOGRAPHY}

We now use the supergravity solutions \eqref{zUpper} and \eqref{zLower} presented in the previous section to calculate the $S^3$ free energy and match the field theory result \eqref{FreeEnergy}. 

\subsection{Supersymmetric holographic renormalization}

In the discussion thus far we have neglected  several issues concerning the behavior of the bulk fields at the $AdS_4$ boundary (in Euclidean signature the $\HH^4$ boundary)  and their effect on the on-shell action. One well-known issue is that
the Euclidean bulk action integral \eqref{ActionEuclidean} diverges at the boundary when classical solutions of the equations of motion are inserted.
The cure for this problem is to introduce a cutoff surface at large distance and add counter-terms to make the action finite.  The theory of holographic renormalization \cite{Skenderis:2002wp} provides a systematic prescription for these counter-terms, but the possibility of finite counter-terms is left open. Finite counter-terms and related issues  are especially important in our problem because the  classical BPS gravity solution  must be dual to the deformed ABJM theory, which possesses global $\cn=2$ supersymmetry on $S^3$.  

It is an axiom of the AdS/CFT correspondence that the classical solutions of the gravity  theory  provide  sources and expectation values for operators in the dual boundary QFT\@.  This information is contained in the asymptotic behavior of bulk fields. To be more specific we consider the $\HH^4$ metric in the form
 \es{H4Hyp}{
  ds^2 = \frac{4 L^2}{\left(1 - r^2 \right)^2} \left( dr^2  +  r^2 d\Omega_3^2 \right)
   = L^2 \left(d\rho^2 + e^{2A(\rho)}\, d\Omega_3^2 \right) \,.
 }
We have made the change of coordinates $r = \tanh (\rho/2)$, so that  $e^{2A(\rho)}= \sinh^2\r$.  We will introduce below a similar radial  coordinate for the metric \reef{GotMetric} of our bulk solution.  In the $\rho$ coordinate,  any solution of the bulk equations of motion of the massless scalars of the gravity theory behaves near the boundary as
 \es{zAsymp}{
  z^\alpha(\rho, x) &= a^\alpha(x) e^{- \rho} + b^\alpha(x) e^{-2 \rho} + \ldots \,, \\
  \tilde z^\alpha(\rho, x) &= \tilde a^\alpha(x) e^{- \rho} + \tilde b^\alpha(x) e^{-2 \rho} + \ldots \,,
 }
where $x$ denotes the coordinates on $S^3$.  In simpler applications of AdS/CFT, the leading coefficient $a^\a(x)$ is the source for the dual operator in the field theory  while $b^\a(x)$ determines the expectation value.  However, for massless scalars this assignment is ambiguous, and we will use global supersymmetry to determine the correct choice.

As originally noted in \cite{Breitenlohner:1982jf}, it turns out that supersymmetry requires the use of regular boundary conditions for some scalars and irregular boundary conditions for others.  When irregular boundary conditions are needed,  the AdS/CFT dictionary requires \cite{Klebanov:1999tb} that a Legendre transformation of the on-shell action be used to calculate quantities in the dual QFT.

In our problem, since we now introduce a spacetime boundary, we should supplement the Einstein-Hilbert action by the well-known Gibbons-Hawking term
 \es{GH}{
  S_\text{GH} = -\frac{1}{8 \pi G_4} \int_{\partial} d^3 x \, \sqrt{h} K \,,
 }
where $h$ is the determinant of the induced metric  $h_{ij}(x, \r)$ at the cutoff $\r = \r_\text{max}$, and $K = \frac 1L \pa_\r \ln \sqrt{h}$ is the trace of the extrinsic curvature.  This boundary term is needed in order to properly define the variational problem for the metric.  Holographic renormalization prescribes that two additional counterterms should be added to the classical action:  a counterterm associated with the curved boundary surface, and a counterterm required because massless scalars are present:
\bea   \lab{cts} 
S_a  &=& \frac{L}{16 \pi G_4} \int_\partial d^3x\, \sqrt{h} {\cal R} \,,\\
S_b   &=& \frac{1}{4 \pi G_4 L} \int_{\partial} d^3 x\, \sqrt{h} \left[1 +\frac 12  \sum_{\alpha=1}^3 z^\alpha \tilde z^\alpha  \right] \,,
\eea   
where ${\cal R}$ is the Ricci scalar of $h_{ij}$. For our situation \eqref{H4Hyp}, the metric diverges at the rate $L^2 e^{2\r}/4$ as $\r \to \infty$ and ${\cal R} =24 e^{-2\r}/L^2$.    As required by holographic renormalization, these counterterms are local functionals of the fields at the cutoff surface.  The sum of the bulk action
in \reef{ActionEuclidean}, with radial integral cut off at $\r=\r_{\rm max}$ and the boundary terms, 
\es{STotal}{
  S = S_\text{bulk} + S_\text{GH} + S_a + S_b \,,
 }
remains finite as $\r_{\rm max}\to\infty$ for any solution of the Euler-Lagrange equations of motion of the gravity theory.  Indeed, the counterterms $S_a$ and $S_b$ are obtained by implementing this requirement, a procedure known as ``near-boundary analysis.''  See \cite{Bianchi:2001kw} or Section~23.11 of \cite{Freedman:2012zz}.

Although well defined, the action $S$ of \reef{STotal} is not satisfactory unless  the counterterm $S_b$ is replaced by
\be
S_{\rm SUSY} =  \frac{1}{4 \pi G_4} \int_{\partial} d^3 x\, \sqrt{h}\, e^{{\cal K}/2} |W| \,.
\ee
The reason for this change is discussed in Appendix~\ref{BOGOMOLNY}.  It is needed to satisfy
global supersymmetry for flat-sliced domain walls, and it is required here as well because both flat- and $S^3$-sliced domain walls are solutions of the same classical theory.  The renormalized on-shell action is then defined as
\be \lab{sren}
S_{\rm on-shell} = S_{\rm bulk} + S_\text{GH} + S_a  +S_{\rm SUSY}\,.
\ee

Inserting the specific K\"ahler potential and superpotential \reef{Superpot}  and  expanding at small $z^\alpha$ and $\tilde z^\alpha$, we rewrite $S_{\rm SUSY} $ as 
\be
S_{\rm SUSY} =  \frac{1}{4 \pi G_4 L} \int_{\partial} d^3 x\, \sqrt{h} \left[1 +\frac 12  \sum_{\alpha=1}^3 z^\alpha \tilde z^\alpha 
   + \frac 12 \left( z^1 z^2 z^3 + \tilde z^1 \tilde z^2 \tilde z^3 \right) \right]\,,
\ee
in which terms which vanish as $\r_{\rm max}\to\infty$ have been dropped. We see that $S_b$ and $S_{\rm SUSY}$ differ by the cubic term in the scalars. The difference makes a finite contribution to $S_{\rm ren}$ in the limit $\r_{\rm max} \to \infty.$  The AdS/CFT match of the free energy, toward which we are working in this section,  depends crucially on the inclusion of $S_{\rm SUSY}.$

%
%%%%%%%%%%%%%%%%%%%%%%%%

A second argument in favor of the counterterm $S_{\rm SUSY}$ emerges from a study of supersymmetric boundary conditions.  At large values of $\r$,  the SUSY transformations of the bulk supergravity theory relate the asymptotic coefficients for the scalar fields $z^\a,\,\, \tz^\a$ in \reef{zAsymp} to the analogous coefficients for the spinors and to each other.
As shown in Appendix C, bulk supersymmetry transformations relate the combinations
 \es{Group1}{
  a^\alpha - \tilde a^\alpha \qquad \text{and} \qquad \left(b^\alpha - \frac{\tilde a^1 \tilde a^2 \tilde a^3}{\tilde a^\alpha}\right)
   + \left(\tilde b^\alpha - \frac{a^1 a^2 a^3}{a^\alpha}\right)
 } 
(separately for each $\a=1,2,3$)  as well as the sets
 \es{Group2}{
  a^\alpha + \tilde a^\alpha \qquad \text{and} \qquad \left(b^\alpha - \frac{\tilde a^1 \tilde a^2 \tilde a^3}{\tilde a^\alpha}\right)
   - \left(\tilde b^\alpha - \frac{a^1 a^2 a^3}{a^\alpha}\right)\,.
 } 
From the viewpoint of the AdS/CFT dictionary,  the boundary behavior \eqref{zAsymp} encodes sources and VEVs in the dual field theory.  
A supersymmetric treatment in the framework of holography requires that we should either take the quantities in \eqref{Group1} to be sources and those in \eqref{Group2} to be VEVs, or the other way around.\footnote{A third possibility exists where we take a linear combination of \eqref{Group1} and \eqref{Group2} to be the sources and the orthogonal linear combination to be the VEVs.  All these possibilities are related by chiral rotations in the bulk.}   Since sources and VEVs are canonically conjugate variables, we would therefore like the canonical conjugate of $a^\alpha$ to be proportional to $\tilde b^\alpha - a^1 a^2 a^3 / a^\alpha$ and that of $\tilde a^\alpha$ to be proportional to $b^\alpha - \tilde a^1 \tilde a^2 \tilde a^3/\tilde a^\alpha$.  This is what the renormalized action \eqref{sren} accomplishes:
 \es{CanConj}{
   \frac{\delta S_\text{on-shell}}{\delta a^\alpha} &= \lim_{\rho \to \infty} e^{-\rho} \left[ 
     \frac{\partial{\cal L}_\text{bulk}}{\partial (\partial_\rho z^\alpha)}
     + \frac{\delta S_\text{SUSY}}{\delta z^\alpha} \right] =  \frac{\sqrt{g^s}L^2}{64 \pi G_4}  \left[ \tilde b^\alpha - a^1 a^2 a^3 / a^\alpha \right] \,, \\
    \frac{\delta S_\text{on-shell}}{\delta \tilde a^\alpha} &= \lim_{\rho \to \infty} e^{-\rho} \left[ 
     \frac{\partial{\cal L}_\text{bulk}}{\partial (\partial_\rho \tilde z^\alpha)}
     + \frac{\delta S_\text{SUSY}}{\delta \tilde z^\alpha} \right] =  \frac{\sqrt{g^s} L^2}{64 \pi G_4}  \left[ b^\alpha - \tilde a^1 \tilde a^2 \tilde a^3/\tilde a^\alpha \right]  \,,
 }
where we denoted the determinant of the metric $g^s_{ij}$ on the unit $S^3$ by $g^s$.  Any additional finite counter-terms cubic in $z^\alpha$ and $\tilde z^\alpha$ would generate additional terms in \eqref{CanConj} that are quadratic in $a^\alpha$ and $\tilde a^\alpha$.  The special role of the counter term $S_{\rm SUSY}$  is thus evident in the result.

\subsection{AdS/CFT dictionary}

The discussion at the end of the previous subsection did not make any assumptions as to whether the coefficients in \eqref{Group1} or those in \eqref{Group2} should be treated as field theory sources or VEVs\@.  From now on we do make such a choice:  We consider the coefficients in \eqref{Group1} to correspond to field theory sources and those in \eqref{Group2} to correspond to VEVs for the following reason.  In Lorentzian signature where $z^\alpha$ and $\tilde z^\alpha$ are complex conjugates of each other, our convention was that the real part of $z^\alpha$ is a scalar and the its imaginary part is a pseudo-scalar.  The bulk scalars are dual to the boundary scalar operators ${\cal O}_B^\alpha$ of scale dimension one, and the bulk pseudo-scalars are dual to the boundary pseudo-scalar operators ${\cal O}_F^\alpha$ of scaling dimension two.  According to the standard rules of the AdS/CFT dictionary, the sources for ${\cal O}_B^\alpha$ and ${\cal O}_F^\alpha$ would then be proportional to $b^\alpha + \tilde b^\alpha$ and $a^\alpha - \tilde a^\alpha$, respectively, at least when all the sources are small.  We interpret the fact that $b^\alpha + \tilde b^\alpha$ should be modified to the expression in \eqref{Group1} as a non-linear effect required by supersymmetry.

In order to read off $(a^\alpha, \tilde a^\alpha, b^\alpha, \tilde b^\alpha)$ more easily, let us now change coordinates  in our solutions \eqref{zUpper} and \eqref{zLower} to the gauge $B=0$ (see \eqref{MetricAnsatz}) where the metric takes the asymptotic form in \eqref{H4Hyp}.  There is no simple analytic formula for this change of coordinates, but we can use the following asymptotic expansion close to the boundary
 \es{rTorho}{
  r = 1 - 2 e^{- \rho} + 2 e^{-2\rho} - 2 \frac{(1 - c_1 c_2 c_3)^2}{(1 + c_1 c_2 c_3)^2} e^{-3 \rho} + \ldots \,,
 }
such that the metric \eqref{GotMetric} takes the form 
 \es{MetricAsymp}{
  ds^2 = L^2 d\rho^2 + \frac {L^2 e^{2 \rho}}4 \left(1 - \frac{1 + c_1 c_2 c_3 (c_1 c_2 c_3 - 10)}{(1 + c_1 c_2 c_3)^2} e^{-2 \rho} + \cdots \right)^2 d\Omega_3^2 \,.
 } 
For the $SU(2)_\ell \times OSp(2 | 2)_r$ branch of solutions \eqref{zUpper} we then have
 \es{zUpperAsymp}{
  z^\alpha(\rho) &= \frac{4 c_\alpha}{1 + c_1 c_2 c_3} e^{-\rho} - \frac{8 c_\alpha (1- c_1 c_2 c_3)}{(1 + c_1 c_2 c_3)^2} e^{-2 \rho} + \cdots \,, \\
  \tilde z^\alpha(\rho) &=  -\frac{4 c_1 c_2 c_3}{c_\alpha (1 + c_1 c_2 c_3)} e^{-\rho} 
     + \frac{8 c_1 c_2 c_3 (1- c_1 c_2 c_3)}{c_\alpha(1 + c_1 c_2 c_3)^2} e^{-2 \rho}+ \cdots \,,
 }
while for the $OSp(2|2)_\ell \times SU(2)_r$ branch the expressions for $z^\alpha$ and $\tilde z^\alpha$ in \eqref{zUpperAsymp} are interchanged.  By comparison of the asymptotic forms \eqref{zUpperAsymp} and \eqref{zAsymp} it is easy to read off the values of $(a^\alpha, \tilde a^\alpha, b^\alpha, \tilde b^\alpha)$.  We can then calculate the values of the sources in \eqref{Group1}:
 \es{Sources}{
  \frac 1a \left( a^\alpha - \tilde a^\alpha \right)  &= \pm \frac{4 \left( c_\alpha + c_1 c_2 c_3 / c_\alpha \right)}{a(1 + c_1 c_2 c_3)}   \,, \\
  \frac 1{a^2} \left[ \left(b^\alpha - \frac{\tilde a^1 \tilde a^2 \tilde a^3}{\tilde a^\alpha}\right)
   + \left(\tilde b^\alpha - \frac{a^1 a^2 a^3}{a^\alpha}\right) \right]
    &=  -\frac{8 \left( c_\alpha + c_1 c_2 c_3 / c_\alpha \right)}{a^2 (1 + c_1 c_2 c_3)}   \,,
 }
where the upper sign corresponds to the branch \eqref{zUpper} and the lower sign corresponds to \eqref{zLower}.  Here, we reintroduced the sphere radius $a$ using dimensional analysis:  the boundary field theory pseudo-scalars sourced by the first line in \eqref{Sources} have scaling dimension two, while the scalars sourced by the second line in \eqref{Sources} have scaling dimension one.    The simplification in the second line of \eqref{Sources} obtained after the inclusion of the quadratic terms in $a^\alpha$ and $\tilde a^\alpha$ is quite remarkable.

To compare with the field theory, we should identify the parameters $\delta_\alpha$ that are related to the R-charges of the bifundamental fields through \eqref{RGeneral} as
 \es{deltaalpha}{
  \delta_\alpha = n \frac{c_\alpha + c_1 c_2 c_3 / c_\alpha}{1 + c_1 c_2 c_3}\,,
 }
where $n$ is a so-far undetermined normalization constant.  Such undetermined normalization constants usually appear in the AdS/CFT dictionary because in most cases there is no clear way of relating the normalization of the bulk field to that of the dual operator.  As we will see below, the supergravity backgrounds in this paper allow the determination of $n$.

Note that, up to a dimensionless factor, we can interpret the quantities $\frac 1a \left( a^\alpha - \tilde a^\alpha\right) $ and $\frac{1}{a^2} \left( b^\alpha - \tilde a^1 \tilde a^2 \tilde a^3 / \tilde a^\alpha + \tilde b^\alpha - a^1 a^2 a^3 / a^\alpha\right)$ as background values for the scalar fields $\sigma'^\alpha$ and $D'^\alpha$ that are part of three (off-shell) background vector multiplets coupled to the boundary theory that we defined in section~\ref{VECTOR}.  In order to preserve SUSY on the boundary we must have\footnote{In \eqref{ThreeBackground} we worked in the case where the SUSY generators are part of $OSp(2|2)_r$.  The $OSp(2|2)_\ell$ case is obtained by sending $a \to -a$, as mentioned in footnote~\ref{ReflectionFootnote}.} $\sigma'^\alpha = \pm i a D'^\alpha$.   Using \eqref{ThreeBackground} and \eqref{deltaalpha}, we find
 \es{GotSigmaD}{
  \sigma'^\alpha = \mp i n \frac{c_\alpha + c_1 c_2 c_3 / c_\alpha}{a (1 + c_1 c_2 c_3)} \,, \qquad
   D'^\alpha = -n \frac{c_\alpha + c_1 c_2 c_3 / c_\alpha}{a^2 (1 + c_1 c_2 c_3) } \,.
 }
These expressions agree indeed with \eqref{Sources} up to a proportionality constant.

\subsection{The $S^3$ free energy}

Evaluating the action~\eqref{sren} on the solutions \eqref{zUpper} and \eqref{zLower}, one obtains
 \es{IOnShell}{
  I = \frac{\pi L^2}{2 G_4} \frac{1 - c_1 c_2 c_3}{1 + c_1 c_2 c_3} \,.
 }
The finite counter-term $S_{\rm SUSY}$ that we included in \eqref{sren} is crucial for obtaining this expression;  without it, the second fraction in \eqref{IOnShell} would be raised to the third power.

It is not $I$ that one should compare with the field theory $F$ because the choice of sources in \eqref{Group1} or \eqref{Group2} involves using the alternate quantization introduced in \cite{Klebanov:1999tb} where for some fields one does not pick the leading coefficient at the boundary to correspond to a source, but instead its canonically conjugate variable.  The quantity $I$ would be the $S^3$ free energy of the field theory where the sources are $a^\alpha$ and $\tilde a^\alpha$.  Instead, we want the $S^3$ free energy in the theory where the sources are \eqref{Group1}, which is just the Legendre transform of the on-shell action with respect to $a^\alpha + \tilde a^\alpha$.  We therefore have
 \es{JDef}{
  J = S_\text{on-shell} - \frac{1}{2} \sum_{\alpha = 1}^3  \int_{S^3}  d^3 x \, \left(a^\alpha + \tilde a^\alpha \right) 
    \left(\frac{\delta S_\text{on-shell}}{\delta a^\alpha} + \frac{\delta S_\text{on-shell}}{\delta \tilde a^\alpha} \right) \,.
 }

For our solutions, using \eqref{CanConj} and the asymptotics \eqref{zAsymp} and \eqref{zUpperAsymp}, we get
 \es{JFinal}{
  J = \frac{\pi L^2}{2 G_4} \frac{\left(1 - c_1^2 \right) \left(1 - c_2^2 \right) \left(1 - c_3^2 \right) }{\left(1 + c_1 c_2 c_3\right)^2} \,.
 }
The pre-factor $\pi L^2 / 2 G_4$ in this expression equals the value of $F$ for the superconformal ABJM theory at level $k=1$ in the large $N$ limit, $\sqrt{2} \pi N^{3/2} / 3$ \cite{Drukker:2010nc, Herzog:2010hf}.  Using this identification, as well as the relation between the R-charges of the bifundamental fields and the $c_\alpha$ as given by \eqref{deltaalpha} and \eqref{RGeneral}, one finds that \eqref{JFinal} agrees with the field theory result \eqref{FreeEnergy} for $n = 1/2$.  This match is one of the main results of this paper.  It is remarkable how when one uses \eqref{deltaalpha} with $n=1/2$, the quantity under the square root in \eqref{FreeEnergy} becomes a perfect square and \eqref{FreeEnergy} matches the expression  \eqref{JFinal}, which has no square roots.

\section{Discussion}

In this paper we found new analytical backgrounds of ${\cal N} = 8$ gauged supergravity in four Euclidean dimensions by solving the BPS equations for an ${\cal N} = 1$ truncation of the ${\cal N} = 8$ theory whose bosonic part consists of the metric and three complex scalar fields.  These solutions, which are given in \eqref{zUpper} and \eqref{zLower}, are asymptotically $\HH^4$, and when embedded into the ${\cal N} = 8$ theory they generically preserve ${\cal N} = 2$ supersymmetry.  They are dual to deformations of the superconformal ABJM theory at Chern-Simons level $k=1$ on $S^3$ corresponding to the most general choice of the ${\cal N} = 2$ $U(1)_R$ symmetry.  Equivalently, they correspond to the ways of coupling the $k=1$ ABJM theory, seen as an ${\cal N} = 2$ theory, to curvature while preserving an $OSp(2|2) \times SU(2)$ symmetry.  On the field theory side, the $S^3$ free energy $F$ was computed in \cite{Jafferis:2011zi} starting from the supersymmetric localization results of \cite{Jafferis:2010un} and using the matrix model techniques developed in \cite{Herzog:2010hf}.  We match this result with a supergravity calculation, as we show in section~\ref{HOLOGRAPHY}.  In obtaining this match, it is important that we perform holographic renormalization in a way consistent with supersymmetry, and that we take a Legendre transform of the on-shell supergravity action.  Our computation provides a way of finding the precise normalization of the field theory operators corresponding to the bulk supergravity scalar fields in our setup. 

It is perhaps worth describing the simplest supergravity backgrounds we find.  If we take $c_2 = c_3 = 0$ in \eqref{zUpper} and \eqref{zLower}, we find that only $z^1$ (in \eqref{zUpper}) or $\tilde z^1$ (in \eqref{zLower}) do not vanish.  These solutions are particularly simple because the scalar fields do not back-react on the metric.  The absence of the back-reaction is due to the fact that the stress tensor involves products of $z^1$ and $\tilde z^1$ (or of their derivatives), as obtained by continuing to Euclidean signature the Lorentzian stress tensor in which $z^1$ and $\tilde z^1$ are each other's complex conjugates;  since either $z^1$ or $\tilde z^1$ vanishes, the stress tensor vanishes too, and by the Einstein equations the metric is just $\HH^4$.    More specifically, the case where $\tilde z^1 = 0$ is
 \es{zUpperSimple}{ 
    ds^2 &= \frac{4 L^2 }{(1-r^2)^2} \left( dr^2 + r^2 d\Omega_3^2 \right) \,, \\
    z^1 &= c_1 (1 - r^2) \,, \qquad
      \tilde z^1 = z^2 = \tilde z^2 = z^3 = \tilde z^3 = 0 \,.
 }
Even though the metric is $\HH^4$, there is no conformal symmetry because $z^1$ depends non-trivially on the $\HH^4$ coordinates.  In Lorentzian signature, a situation where the metric is $AdS_4$ but the matter fields break the $SO(3, 2)$ symmetry would be impossible, because any complex scalar with a non-trivial profile in the $AdS_4$ directions produces a non-vanishing stress tensor.

For a superconformal field theory dual to $AdS_4 \times Y$, where $Y$ is a seven-dimensional Sasaki-Einstein space, it was shown in \cite{Herzog:2010hf} that the $S^3$ free energy is given by
 \es{FSE}{
  F = N^{3/2} \sqrt{\frac{2 \pi^6}{27 \Vol(Y)}} \,.
 }
For ABJM theory at level $k=1$, we have $Y = S^7$.  Using $\Vol(S^7) = \pi^4/3$, it is not hard to see that \eqref{FSE} matches \eqref{FreeEnergy} when the R-charges have the superconformal values $R[Z^a] = R[W_b] = 1/2$.  An interesting observation of \cite{Jafferis:2011zi, Martelli:2011qj} was that, in fact, \eqref{FSE} continues to hold away from the superconformal values of the R-charges provided that one computes $\Vol(Y)$ using a Sasakian metric on $Y$ that is not Einstein.  The volumes of these Sasakian metrics can be taken to be parameterized by the R-charges $R[Z^a]$ and $R[W_b]$, computed now as volumes of certain five-cycles of $Y$.  There is so far no known explanation of this observation.  It would be interesting to see whether the Euclidean supergravity solutions constructed in this paper can be lifted to eleven dimensions, and whether the lift would illuminate why \eqref{FSE} still holds away from the superconformal point.

As is well known, the gauge/gravity duality provides a valid description of the dual quantum field theory in a strong coupling limit, a limit in which traditional field-theoretic methods usually cannot be applied. In the example considered in this paper, the field theory results of \cite{Jafferis:2011zi} were made possible by the method of supersymmetric localization, which was quite powerful even at strong coupling and in the absence of conformal symmetry.  Thus the agreement  we have found  between the two sides of the duality is unusually precise and quantitative.

\section*{Acknowledgments}

We thank H.~Elvang, D.~Jafferis, I.~Klebanov,  J.~Maldacena, M.~Mezei, and K. Skenderis for useful discussions.  The research
of DZF is supported in part by NSF grant PHY-0967299.
The work of SSP is supported in part by a Pappalardo Fellowship in Physics at MIT.
Both authors are supported in part by the U.S. Department of Energy under cooperative research agreement Contract Number DE-FG02-05ER41360\@.  SSP thanks the Stanford Institute for Theoretical Physics  for hospitality while this work was in progress.

\appendix

\section{Conventions and Euclidean supersymmetry}
\label{CONVENTIONS}

In this Appendix we explain our conventions, and comment on the Euclidean continuation of supersymmetric $(3+1)$-dimensional theories in flat space.  We focus on theories with global supersymmetry because going from global to local supersymmetry in Euclidean signature doesn't pose any additional challenges relative to the ones in Lorentzian signature.  The results presented in this section are not new.  They are implicit, for example, in the recent work \cite{Festuccia:2011ws}.

\subsection{Lorentzian signature conventions}

Our Lorentzian signature conventions are the same as those in \cite{Freedman:2012zz}.   As in Chapter~3 of \cite{Freedman:2012zz}, we define the four-dimensional gamma matrices to be
 \es{gammamu}{
  \gamma^\mu = \begin{pmatrix}
   0 & \sigma^\mu \\
   \bar \sigma^\mu & 0 
  \end{pmatrix}  \,, \qquad 
   \begin{array}{c}
   \sigma^\mu = \left( 1, \vec{\sigma} \right) \,, \\
   \qquad \bar \sigma^\mu = \left( -1, \vec{\sigma} \right) \,,
   \end{array}
 }
where $\sigma^i$ are the three Pauli matrices.  These matrices satisfy the following two useful relations:
 \es{sigmaRel}{
  \gamma^0 \gamma^\mu &= \begin{pmatrix}
   \bar \sigma^\mu & 0 \\
   0 & - \sigma^\mu
  \end{pmatrix}  \,, \qquad \sigma^2 \sigma^\mu \sigma^2 = - (\bar \sigma^\mu)^T \,.
 }
The left and right projectors are $P_L = (1 + \gamma^5 )/ 2$ and $P_R = (1 - \gamma^5 )/2$, where $\gamma^5$ is given by
 \es{gamma5}{
   \gamma^5 = -i \gamma^0 \gamma^1 \gamma^2 \gamma^3 = \begin{pmatrix}
    1 & 0 \\
    0 & -1 
   \end{pmatrix} = \sigma_3 \otimes 1 \,.
 }

In \cite{Freedman:2012zz}, supersymmetric theories in $3+1$ dimensions are constructed using Majorana spinors.\footnote{We treat all spinor fields and parameters as anti-commuting.}  The Majorana condition on a Dirac spinor $\hat \chi$ is $\hat \chi = B^{-1} \hat \chi^*$, where
 \es{BC}{
  B &= \gamma^0 \gamma^1 \gamma^3 = \begin{pmatrix}
   0 & -i \sigma_2 \\
   i \sigma_2 & 0 
  \end{pmatrix} = \sigma_2 \otimes \sigma_2 = B^{-1}\,. % \,, \\
 } 
Writing $\hat \chi$ in Weyl form
 \es{psiWeyl}{
  \hat \chi = \begin{pmatrix}
   \chi \\
   \tilde{\chi}
  \end{pmatrix} \,,
 }
the Majorana condition implies
 \es{chietaMajorana}{
   \tilde{\chi} = i \sigma_2 \chi^* \,,\qquad \chi^\dag = \tilde{\chi}^Ti\s_2 \,.
 }
The Dirac adjoint is defined as $\bar{\hat \chi} = \hat \chi^\dag i \g^0$.  
The result
\be \lab{4to2}
\bar{\hat \chi} = i \begin{pmatrix} -\tilde{\chi}^\dag & \chi^\dag \end{pmatrix} = i \begin{pmatrix} \chi^Ti\s_2 & \tilde{\chi}^Ti\s_2 \end{pmatrix}
\ee
is helpful to convert 4-component expressions to Weyl form.

As an example, the Lagrangian of a massive Majorana fermion can be written in Weyl components as
 \bea \lab{MajLag}
	 -{\cal L}_\text{Majorana} &=& \frac 12 \bar { \hat \chi} \slashed{\partial} \hat \chi - \frac 12 m \bar {\hat \chi} \hat \chi\\
    &=&\frac i2 \tilde \chi^T (i \sigma_2) \bar \sigma^\mu \overleftrightarrow{\partial_\mu} \chi 
     - \frac i2 m \left[\tilde \chi^T (i \sigma_2) \tilde \chi + \chi^T (i \sigma_2) \chi \right]\,.     
\eea

As another example, the gravitino Lagrangian is
 \es{Gravitino}{
  {\cal L}_\text{gravitino} &= - \frac 12 \bar {\hat \psi}_\mu \gamma^{\mu\nu\rho} D_\nu \hat \psi_\rho - \frac{m}{2} \bar {\hat \psi}_\mu \gamma^{\mu\nu} \hat \psi_\nu \\ 
   &= \frac i2 \epsilon^{\mu\nu\rho\sigma} \bar {\hat \psi}_\mu  \gamma_5 \gamma_\nu D_\rho {\hat \psi}_\sigma
    - \frac{m}{2} \bar {\hat \psi}_\mu \gamma^{\mu\nu} {\hat \psi}_\nu \,,
 }
where we used $\gamma^{\mu\nu\rho} = -i \epsilon^{\mu\nu\rho\sigma} \gamma_5 \gamma_\sigma$ and $\epsilon^{0123} = -1$.  The Weyl decomposition is
\es{GravitinoFinal}{
  {\cal L}_\text{gravitino} =  -\frac 12 \epsilon^{\mu\nu \rho \sigma}
   \tilde \psi_\mu^T (i \sigma_2) \bar \sigma_\nu \overleftrightarrow{D_\rho} \psi_\sigma
    - \frac{m}2 \left[  \psi_\mu^T  i \sigma_2 \sigma^{[\mu} \bar \sigma^{\nu]} \psi_\nu + \tilde \psi_\m^T (i \sigma_2) \bar{\s}^{[\mu} \sigma^{\nu]} \tilde \psi_\nu \right]\,.
 }

Note that a feature of this notation (which persists in the Euclidean version below) is that no special concern is needed for up/down or dotted/undotted spinor indices.

%%%%%%%%%%%%%%%%%%%%%%%
\subsection{From Lorentzian to Euclidean signature}

We use a very straightforward method to define the Euclidean version of any 4-dimensional  Lorentzian signature  field theory.  There are three steps:
\begin{enumerate}
\item Rewrite  the Lorentzian theory in terms of two-component Weyl spinors.
This allows us to treat Lorentzian theories with either Dirac or Majorana spinors.
\item Continue the time components of vectors as $x^0\to -ix^4, ~ A_0 \to i A_4,~\s_0 \to i\s_4,~\bar{\s}_0 \to  i\bar{\s}_4.$  After this is done, the Lorentzian and Euclidean actions are related by $\exp \left[ {iS^\text{Lor}}\right]=\exp \left[-S^\text{Euc}\right]$. 
\item Require that the resulting action, transformation rules, and equations of motion are invariant under the spacetime isometry group $SO(4)$, which is implemented  as $SU(2)_L \times SU(2)_R$.  
\end{enumerate}

Euclidean symmetry is the guiding principle.  To discuss its application to spinors we define Euclidean ``Weyl matrices'' by
  \es{eucsigma}{
    \s_\m = (\vec{\s}, - i)\,, \qquad \bar{\s}_\m =(\vec{\s}, i)\,,
 }
where  the $\s^i$ are again the Pauli matrices.\footnote{In flat Euclidean space, there is no distinction between upper and lower vector indices.}  As in the Lorentzian case, the two sets of matrices are related by
\be  \lab{sigtosigbar}
\sigma^2 \sigma_\mu \sigma^2 = - (\bar \sigma_\mu)^T \,.
\ee

Let $y^\m$ be a real 4-vector and define the matrix 
\be 
Y = \bar{\s}_\m y^\m \,,
\ee
which is an imaginary scalar multiple of a finite element of $SU(2)$.   Note that $\det Y = - \sum_\m y^\m y^\m$, and that the vector components can be obtained from $Y$ as the trace
\be  
y^\m = \frac12 \tr (\s^\m Y)\,.
\ee  
Let the pair of  matrices $(U,V)$ denote an element of $SU(2)_L \times SU(2)_R$.  The transformation
\be\lab{homom}
Y \to Y' \equiv V^{-1} Y U
\ee
defines a linear map that takes $Y$ into $Y'$, which is another matrix of the same type. Since the determinant is invariant and $SU(2)$ is a connected group, the vector $y'^{\m} $ is related to $y^\m$ by  $y'^\m = \Lambda^\m{}_\n y^\n$, where $\L$ is a matrix of $SO(4)$.   Indeed, the map \reef{homom} defines a homomorphism\footnote{When $V=U$,  the corresponding $SO(4)$ transformations   fix the component $y^4$ of $y^\mu$,  while transformations with
$V^{-1} =U$ correspond to ``boosts'' involving $y^4$.} of $SU(2)_L \times SU(2)_R \to SO(4).$ 

The Majorana spinor Lagrangian \reef{MajLag} looks essentially the same in Euclidean signature 
\be \lab{MajoranaEuclidean}
   {\cal L}_\text{Majorana} = \frac i2 \tilde \chi^T (i \sigma_2) \bar \sigma^\mu \overleftrightarrow{\partial_\mu} \chi 
     - \frac i2 m \left[\tilde \chi^T (i \sigma_2) \tilde \chi + \chi^T (i \sigma_2) \chi \right]\,,
 \ee
but it is now interpreted  as a Euclidean Lagrangian with  $\bar{\s}^\m$ given in \reef{eucsigma}. The kinetic term is  invariant provided that  spinors transform as 
\be
\chi(x) \to U \chi(\L^{-1}x) \,,  \qquad \tilde \chi(x) \to V \tilde \chi( \L^{-1}x) \,.
\ee
The mass terms in \eqref{MajoranaEuclidean} reduce to the standard $SU(2)$ invariant form, e.g.
\be
 \chi^T i \s_2 \chi =  \eps^{\a\b}\chi_\a\chi_\b\,.
\ee
The reason that $\tilde \chi$ cannot be interpreted as the Hermitian conjugate of $\chi$ now emerges;  they transform in non-conjugate representations of $SO(4)$.  The spinor $\chi$ transforms in the (1/2,0) representation of $SU(2)_L\times SU(2)_R$, and $\tilde \chi$ transforms in the (0,1/2) representation.   Thus we treat $\chi$ and $\tilde \chi$ as independent fields in the Euclidean theory.

The equations of motion which follow from \reef{MajoranaEuclidean} are
 \be \lab{eomsEuclidean}
  \bar \sigma^\mu \partial_\mu \chi = m \tilde \chi \,, \qquad
   \sigma^\mu \partial_\mu \tilde \chi = m \chi \,.
 \ee
To derive the second equation we used \reef{sigtosigbar}.  Combining these two equations we obtain
 \be  \lab{KGeuclidean}
  \nabla^2 \chi = \sigma^\mu \partial_\mu \bar \sigma^\nu \partial_\nu \chi = m^2 \chi \,.
 \ee
The sign of the $m^2$ term is the same as for a massive scalar field in Euclidean space, as it should be.  (Note that the propagators of massive Euclidean fields have no poles.)

 \subsection{Euclidean Supersymmetry}
 \label{EUCLIDEANSUSY}
 
The next step is to study Euclidean supersymmetry. To stimulate book sales
 we start with the component form of the $\cn = 1,~ D=4$ theories  discussed in Chapter 6 of \cite{Freedman:2012zz}.    The conventions there are modified as follows:
\begin{enumerate}
\item   Scale the SUSY parameters of \cite{Freedman:2012zz} by $\e \to \sqrt{2}\e$.
\item   Since there are no Majorana spinors in the Euclidean theory,  the
formulas of \cite{Freedman:2012zz} must be rexpressed in the Weyl spinor formalism using Appendix A above.
\end{enumerate}
We then  use the procedure outlined in Appendix B.  The Euclidean versions of the supersymmetry and supergravity theories needed in the main part of this paper were obtained by this method.

We now give two simple examples that we will also use later on in Appendix~\ref{SUSY3D}.  The first is that of an off-shell $U(1)$ vector multiplet, which in Lorentzian signature consists of a gauge field $A_\mu$, a Majorana fermion $\hat \lambda$, and a real auxiliary scalar $D$.  The Lorentzian Maxwell action is (see (6.48) of~\cite{Freedman:2012zz} with $\lambda \to \hat \lambda / \sqrt{2}$)
 \es{SMaxwellLor}{
  S_\text{Maxwell}^\text{Lor} = \frac{1}{e^2} \int d^4 x\, \left(-\frac{1}{4} F_{\mu\nu} F^{\mu\nu} - \frac 14 \bar {\hat \lambda} \gamma_\mu D^\mu \hat \lambda + \frac 12 D^2 \right) \,.
 }
In Euclidean signature, we formally allow $A_\mu$ and $D$ to be complex, and consider the Weyl components $(\lambda, \tilde \lambda)$ of $\hat \lambda$ (see \eqref{psiWeyl}) to be independent.  The Euclidean Maxwell action is therefore
 \es{SMaxwell}{
  S_\text{Maxwell}^\text{Euc} = \frac{1}{e^2} \int d^4 x\, \left(\frac{1}{4} F_{\mu\nu} F^{\mu\nu} + \frac i4 \tilde \lambda^T (i \sigma_2) \bar \sigma^\mu \overleftrightarrow{\partial_\mu} \lambda - \frac 12 D^2  \right) 
 }
Both \eqref{SMaxwellLor} and \eqref{SMaxwell} are invariant under the transformation rules
 \es{Gauge4dTransf}{
  \delta A_\mu &= - \frac i2 \left[\epsilon^T (i \sigma_2) \sigma_\mu \tilde \lambda + \tilde \epsilon^T (i \sigma_2) \bar \sigma_\mu \lambda \right] \,, \\
  \delta D &= -\frac 12 \left[\epsilon^T (i \sigma_2) \sigma^\mu \partial_\mu \tilde \lambda - \tilde \epsilon^T(i \sigma_2) \bar \sigma^\mu \partial_\mu \lambda \right] \,, \\
  \delta \lambda &= \left(  \frac 12  \sigma^{[\mu} \bar \sigma^{\nu]} F_{\mu\nu} + i D \right) \epsilon \,, \\
  \delta \tilde \lambda &= \left( \frac 12 \bar \sigma^{[\mu} \sigma^{\nu]} F_{\mu\nu} - i D \right) \tilde \epsilon \,,
 }
which can be found by writing the transformation rules given in~(6.49) of~\cite{Freedman:2012zz} in Weyl components.  These transformation rules are correct in both Lorentzian and Euclidean signature provided that one uses the Weyl matrices in \eqref{gammamu} in Lorentzian signature and \eqref{eucsigma} in Euclidean.    The Lorentzian theory requires $\tilde \lambda = (i \sigma_2) \lambda^*$ and $\tilde \epsilon = (i \sigma_2) \epsilon^*$, but in the Euclidean theory we should take $\tilde \lambda$ and $\tilde \epsilon$ to be independent from $\lambda$ and $\epsilon$.

As a second example, we study a massless ${\cal N} = 1$ chiral multiplet charged under a $U(1)$ vector multiplet.  In Lorentzian signature, the chiral multiplet fields are a complex boson $Z$, its complex conjugate $\tilde Z = \bar {Z}$, a Majorana fermion $\hat \chi$, and a complex auxiliary boson $F$ and its complex conjugate $\tilde F = \bar{F}$.  The action is given in (6.58) and (6.59) of \cite{Freedman:2012zz}:
 \es{Chiral4dLor}{
  S_\text{chiral}^\text{Lor} = \int d^4 x\, \left(- D^\mu \bar Z D_\mu Z - \bar{\hat \chi} \gamma^\mu P_L D_\mu \hat \chi + \bar F F 
   + i \bar{\hat \lambda} \bar Z P_L \hat \chi + i \bar {\hat \chi} P_R Z \hat \lambda + D \bar Z Z \right)  \,.
 } 
The covariant derivatives appearing in this expression are 
 \es{CovDers}{
   D_\mu Z &= (\partial_\mu - i A_\mu)Z \,, \\
   D_\mu \bar Z &= (\partial_\mu + i A_\mu) \bar Z \,, \\
   D_\mu P_L \hat \chi &= (\partial_\mu - i A_\mu) P_L \hat \chi \,, \\
   D_\mu P_R \hat \chi &= (\partial_\mu + i A_\mu) P_R \hat \chi \,.
  }
Passing to Weyl components and performing the Euclidean continuation, we obtain the Euclidean action
 \es{Chiral4dEuc}{
  S_\text{chiral}^\text{Euc} = \int d^4 x\, \left(D^\mu \tilde Z D_\mu Z + i \tilde \chi^T (i \sigma_2) \bar \sigma^\mu D_\mu \chi - \tilde F F 
   + \lambda^T (i \sigma_2) \tilde Z \chi + \tilde \chi^T (i \sigma_2) Z \tilde \lambda - D \tilde Z Z \right) \,.
 }
The transformation rules are obtained from (6.62) and (6.63) of \cite{Freedman:2012zz}:
 \es{ChiralTransf4d}{
  \delta Z &= i \epsilon^T (i \sigma_2) \chi \,, \\
  \delta \tilde Z &= i \tilde \epsilon^T (i \sigma_2) \tilde \chi \,, \\
  \delta \chi &= \sigma^\mu D_\mu Z \tilde \epsilon + F \epsilon \,, \\
  \delta \tilde \chi &= \bar \sigma^\mu D_\mu \tilde Z \epsilon + \tilde F \tilde \epsilon \,, \\
  \delta F &= i \tilde \epsilon^T (i \sigma_2) \bar \sigma^\mu D_\mu \chi - \tilde \epsilon^T (i \sigma_2) \tilde \lambda Z \,, \\
  \delta \tilde F &= i \epsilon^T (i \sigma_2) \sigma^\mu D_\mu \tilde \chi + \epsilon^T (i \sigma_2) \lambda \tilde Z \,.
 }
As in the case of the vector multiplet, these transformation rules are equally valid in both Lorentzian and Euclidean signature, provided that one uses \eqref{gammamu} in Lorentzian signature and \eqref{eucsigma} in Euclidean.  Furthermore, in Lorentzian signature we should impose the reality conditions $\tilde Z = \bar Z$, $\tilde F = \bar F$,  $\tilde \chi = (i \sigma_2) \chi^*$, and $\tilde \epsilon = (i \sigma_2) \epsilon^*$, but these reality conditions are formally relaxed in Euclidean signature.

\section{BPS form of flat-sliced domain wall action}
\label{BOGOMOLNY}

In this section we put the action for a flat-sliced AAdS 4-dimensional domain wall into the Bogomolny form of a sum of squares plus surface term.  The  first order differential equations which appear as factors in the square terms then become the BPS equations for the domain wall and the surface term becomes the counter term needed for the on-shell action.
We work in Lorentzian signature.

We write the domain wall metric as
\be\lab{domwall}
ds^2 = e^{2A(r)} \h_{\m\n}dx^\m dx^\n  + dr^2 \,.
\ee
It is straightforward to work out the scalar curvature for an arbitrary bulk dimension $D$:
 \es{ScalarR}{
  R = -(D-1)\left(2A'' + D A'^2 \right) \,.
 }
After partial integration the gravitational action becomes:
\be \lab{sdom}
 \frac{1}{16 \pi G_D} \int \sqrt{-g} R = \frac{(D-1)(D-2)}{16 \pi G_D} \int dr d^{D-1}x\, e^{(D-1)A} A'^2\,.
\ee
The surface term that is dropped vanishes exponentially at the AdS boundary.  If evaluated at a finite cutoff $r=r_{\rm max}$, it is canceled by the Gibbons-Hawking boundary term present in the action.

For a general ${\cal N} = 1$ K\"ahler $\sigma$-model coupled to gravity with potential of the form \eqref{VFGeneral}, we find
\bea \lab{Sbps}  \nonumber
S &=& \frac{1}{8\pi G_4}\int d^4x\sqrt{-g}[ \frac12 R - {\cal K}_{\a\bb}\pa_\m z^\a \pa^\m \bz^{\bb} - V_F]\\  \nonumber
&=& \frac{1}{8\pi G_4}\int dr\, d^3x \left( e^{3A}\left[ 3 (\pa_r A \pm e^{{\cal K}/2} |W|)^2 - {\cal K}_{\a\bb} (\pa_r z^\a \mp e^{{\cal K}/2} \sqrt{W/\bar{W}}\, {\cal K}^{\a\bg}\nabla_{\bg}\bW)\right.\right.\\
&&\quad\left.\left.(\pa_r \bz^{\bb}\mp e^{{\cal K}/2} \sqrt{\bW/W}\, K^{\d\bb}\nabla_\d W) \right] - \frac{\pa}{\pa r}(2e^{3A} e^{{\cal K}/2} |W|)  \right) \,.
\eea
Note that each factored term and the surface term are invariant under K\"ahler transformations ${\cal K}(z,\bz) \to {\cal K}(z,\bz)+ f(z) +\bar{f}(\bz)$.  We see
that the action is stationary if the first order flow equations we now write are satisfied:
 \es{floweqs}{
\pa_r z^\a &=\pm e^{{\cal K}/2} \sqrt{W/\bar{W}}\, {\cal K}^{\a\bg}\nabla_{\bg}\bW \,, \\
 \pa_r \bz^{\bb}&= \pm e^{{\cal K}/2} \sqrt{\bW/W}\, {\cal K}^{\d\bb}\nabla_\d W \,, \\
\pa_r A &= \mp e^{{\cal K}/2} |W| \,.
 }
These are the BPS equations for general flat-sliced domain walls.

On-shell, when the BPS equations are satisfied, the action \reef{Sbps} is given by the total derivative term.   
At the boundary cutoff, we find
\be
 S_\text{cutoff}  = \frac{1}{4\pi G_4}\int d^3x \left(e^{3A} e^{{\cal K}/2} |W|\right)_{r=r_{\rm max}}\,.
 \ee
Since $e^{A(r)} \sim e^{r/L}$  as $r\to \infty$, $L$ being the AdS scale, the surface term is divergent with cubic leading divergence.  We must add a counterterm to cancel the divergence at the boundary in order to construct the renormalized on-shell action.  The formalism of holographic renormalization requires that the counterterm is  a local function of the supergravity fields, evaluated at the cutoff, but this prescription allows unspecified finite local counterterms.  In this case,  we must resolve this ambiguity by choosing the  counterterm $S_\text{ct} = -S_\text{cutoff}$, so that the renormalized on-shell action \emph {of a BPS domain wall} vanishes.  Otherwise the expectation value  $\<T_{\m\n}\>$ of the stress tensor in the state of boundary QFT dual to the bulk BPS configuration would not vanish, in violation of global supersymmetry.  To obtain correlation functions in the boundary field theory, we  need to consider the more general bulk configuration of \emph{BPS domain wall plus fluctuations}.  In this case the renormalized on-shell action does not vanish.
However,  as an example in five bulk dimensions shows \cite{Bianchi:2001de, Bianchi:2001kw}, it is necessary to include the counterterm above,
evaluated in the more general configuration, or otherwise correlation functions may have unphysical properties.

In this paper we are primarily interested in $S^3$-sliced domain walls, rather than flat-sliced.  The Bogomolny argument fails in this case, and the renormalized action of the domain wall configuration does not vanish.
However,  the counterterms of holographic renormalization have a ``universal'' structure. They must be valid for \emph{all} solutions of the classical field equations of a given bulk theory.  Therefore we include the same counterterm above in the calculation of the on-shell action in Section~\ref{HOLOGRAPHY}.

\section{The boundary limit of bulk SUSY}

The main goal of this Appendix is to understand how bulk SUSY transformations act on the coefficients $(a^\alpha, \tilde a^\alpha, b^\alpha, \tilde b^\alpha)$ defined through \eqref{zAsymp}.  A similar analysis was performed in \cite{Amsel:2008iz} in a slightly simpler situation in Lorentzian signature.

It can be checked that the coefficients $(a^\alpha, \tilde a^\alpha, b^\alpha, \tilde b^\alpha)$ mix only with the  coefficients of the leading asymptotic terms $e^{-3\rho/2}$ and $e^{-5\rho/2}$ in the expansions of the bulk gauginos $\chi^\alpha$ and $\tilde \chi^\alpha$.   In particular, they don't mix with the gravitino and they don't back-react on the metric.  We therefore work with the $\HH^4$ metric given in \eqref{H4Hyp} (with radial coordinate $\r$).

The bosonic action \eqref{ActionEuclidean} must be supplemented by its fermionic counterpart.  The full action can be found in Chapter~18.1 of \cite{Freedman:2012zz} and is fairly complicated.  Only a limited number of fermionic terms, namely the kinetic and mass terms for the bulk gauginos, are relevant for the near boundary analysis we are about to perform, the rest being suppressed.  In Lorentzian signature, the quadratic action for the bulk gauginos is in four-component notation 
 \es{DilatQuadLor}{
  {\cal L}_f = -\frac12 {\cal K}_{\alpha \bb} \left( \bar {\hat \chi}^\alpha \slashed{D} P_R \hat \chi^{\bb} 
   + \bar {\hat \chi}^{\bb} \slashed{D} P_L\hat \chi^{\alpha} \right) 
   - \frac 12 \left(m_{\alpha\beta} \bar {\hat \chi}^\alpha P_L \hat \chi^\beta
    + \bar m_{\bar \alpha \bb}  \bar {\hat \chi}^{\bb} P_R \hat \chi^{\bar \alpha} \right) \,.
 }
The K\"ahler metric ${\cal K}_{\alpha \bar \beta}$ and other target space data we need are given in \eqref{Kahler}--\eqref{nablaW}.  The mass matrices can be obtained from the superpotential via
 \es{DilatinoMass}{
  m_{\alpha\beta} = e^{{\cal K}/2} \nabla_\alpha \nabla_\beta W \,, \qquad
   \bar m_{\bar \alpha \bb} = e^{{\cal K}/2} \bar \nabla_{\bar \alpha} \bar \nabla_{\bar \beta} {\bar W} \,.
 }
Since we are only interested in the asymptotic expansion at large $\rho$ where the scalars are small, we can expand both the K\"ahler metric and the mass matrices at small values of the fields.  Using the K\"ahler metric \eqref{Kahler} and the superpotential \eqref{Superpot}, we find
 \es{Expansions}{
  {\cal K}_{\alpha \bar \beta} &= \delta_{\alpha \bar \beta} + \cdots \,, \\
  m_{\alpha \beta} &= \frac{z^1 z^2 z^3}{z^\alpha z^\beta} + \cdots \,, \qquad
   \bar m_{ \bar \alpha \bb} =  \frac{\bar z^{1} \bar z^{2} \bar z^{3}}{\bar z^{\bar \alpha} \bar z^{\bar \beta}} + \cdots \,.
 }

The equations of motion following from \eqref{DilatQuadLor} are
 \es{eomsDilatinos}{
  {\cal K}_{\alpha \bb} \slashed{D} (P_L \hat \chi^\alpha)  &= - \bar m_{\bar \alpha \bb} (P_R \hat \chi^{\bar \alpha}) \,, \\
  {\cal K}_{\alpha \bb} \slashed{D} (P_R \hat \chi^{\bb}) &= - m_{\alpha \beta} (P_L \hat \chi^\beta) \,.
 }
These equations hold both in Lorentzian and Euclidean signature, provided that one uses the appropriate form of the gamma matrices.   For the Euclidean case where the metric is \eqref{H4Hyp}, expanding these equations at large $\rho$, we find
 \es{chiExpansion}{
  \chi^1(\rho, x) &= \gamma^1(x) e^{-3 \rho / 2} - i\left( 2 \slashed{\nabla} \gamma^1(x) 
     + \tilde a^2(x) \tilde \gamma^3(x) + \tilde a^3(x) \tilde \gamma^2(x) \right) e^{-5 \rho/2} + \ldots \,, \\
  \tilde \chi^1(\rho, x) &= \tilde \gamma^1(x) e^{-3 \rho / 2} +  i \left(2 \slashed{\nabla} \tilde \gamma^1(x) 
     + a^2(x) \gamma^3(x) + a^3(x) \gamma^2(x) \right) e^{-5 \rho/2} + \ldots \,,
 }
as well as similar relations obtained by cyclic permutations of the $123$ indices. Here, $\slashed{\nabla}$ is the Dirac operator on the unit $S^3$, and $\gamma^\alpha(x)$ and $\tilde \gamma^\alpha(x)$ are spinor-valued integration constants that depend only on the coordinates on $S^3$.  These equations are the fermionic analog of \eqref{zAsymp}.

Bulk supersymmetry transformations will mix together $(a^\alpha, \tilde a^\alpha, b^\alpha, \tilde b^\alpha, \gamma^\alpha, \tilde \gamma^\alpha)$.  The expressions for the supersymmetry variations of the bulk gauginos were given in \eqref{SUSYvariationsEuclidean}.  They should be supplemented by the supersymmetry variations of the bosons:
 \es{zSUSYvar}{
  \delta z = i \chi^T (i \sigma_2) \epsilon \,, \\
  \delta \tilde z = i \tilde \chi^T (i \sigma_2) \tilde \epsilon \,.
 }

The boundary expansion of the bulk Killing spinors in the case where they are proportional to the left-invariant $S^3$ Killing spinors $\zeta$ is
 \es{KillingAsymp}{
  \epsilon(\rho, x) &= \left(  e^{\rho/2}  - e^{-\rho/2} + \cdots \right) \zeta(x) \,, \\
  \tilde \epsilon (\rho, x) &= -i \left(e^{\rho/2}  + e^{-\rho/2} + \cdots  \right) \zeta(x) \,.
 }
This expression follows from \reef{rTorho} combined with either \reef{KillingH4Soln1}
or \reef{KillingUpper} (up to an unimportant normalization constant).  Using the transformation rules and \reef{KillingAsymp}, we find
 \es{SUSYvars}{
  \delta a^1 &= i \gamma^{1T} (i \sigma_2) \zeta \,, \\
  \delta \tilde a^1 &= \tilde \gamma^{1T} (i \sigma_2) \zeta \,, \\
  \delta b^1 &= \left[2 \slashed{\nabla} \gamma^1 + (\tilde a^2 \tilde \gamma^3 + \tilde a^3 \tilde \gamma^2) - i \gamma^1 \right]^T (i \sigma_2) \zeta \,, \\
  \delta \tilde b^1 &= i \left[2 \slashed{\nabla} \tilde \gamma^1 + (a^2 \gamma^3 + a^3 \gamma^2) - i \tilde \gamma^1 \right]^T (i \sigma_2) \zeta \,, \\
  \delta \gamma^1 &= \left(2 a^1 + b^1 - 2 i \slashed{\partial} a^1 - \tilde a^2 \tilde a^3 \right) \zeta \,, \\
  \delta \tilde \gamma^1 &= i \left(2 \tilde a^1 - \tilde b^1 - 2 i \slashed{\partial} \tilde a^1 + a^2 a^3 \right) \zeta \,.
 }
These are the sought for asymptotic transformation rules!  They can be regrouped into the form 
 \es{SUSYvarsBetter}{
  \delta (a^1 \mp \tilde a^1) &= i \left(\gamma^1 \pm i \tilde \gamma^1 \right) \zeta \,, \\
  \delta \left((b^1 - \tilde a^2 \tilde a^3) \pm (\tilde b^1 - a^2 a^3) \right) &= 
   \left[2 \slashed{\nabla} \left(\gamma^1 \pm i \tilde \gamma^1 \right) - i \left(\gamma^1 \pm i \tilde \gamma^1 \right)  \right]^T (i \sigma_2) \zeta \,, \\
   \delta \left(\gamma^1 \pm i \tilde \gamma^1 \right)  &= \left(2 (a^1 \mp \tilde a^1) - 2 i \slashed{\partial} (a^1 \mp \tilde a^1)
    +  (b^1 - \tilde a^2 \tilde a^3) \pm (\tilde b^1 - a^2 a^3) \right) \zeta \,,
 }
 which shows that the asymptotic constants of the bulk fields split into two sets (corresponding to the upper and lower signs) by the action of the SUSY transformations.  We take the combinations with the upper signs to correspond to field theory sources and the ones with lower signs to correspond to VEVs.

A similar analysis can be performed for the case where the bulk Killing spinors are proportional to the right-invariant $S^3$ Killing spinors $\xi$.  We have
 \es{KillingAsymp2}{
  \epsilon(\rho, x) &= \left(  e^{\rho/2}  + e^{-\rho/2} + \cdots \right) \xi(x) \,, \\
  \tilde \epsilon (\rho, x) &= -i \left(e^{\rho/2}  - e^{-\rho/2} + \cdots  \right) \xi(x) \,,
 }
which up to normalization follows from a large $\rho$ expansion of \eqref{KillingLower}.  We find
 \es{SUSYvars2}{
  \delta a^1 &= i \gamma^{1T} (i \sigma_2) \xi \,, \\
  \delta \tilde a^1 &= \tilde \gamma^{1T} (i \sigma_2) \xi \,, \\
  \delta b^1 &= \left[2 \slashed{\nabla} \gamma^1 + (\tilde a^2 \tilde \gamma^3 + \tilde a^3 \tilde \gamma^2) + i \gamma^1 \right]^T (i \sigma_2) \xi \,, \\
  \delta \tilde b^1 &= i \left[2 \slashed{\nabla} \tilde \gamma^1 + (a^2 \gamma^3 + a^3 \gamma^2) + i \tilde \gamma^1 \right]^T (i \sigma_2) \xi \,, \\
  \delta \gamma^1 &= \left(-2 a^1 + b^1 - 2 i \slashed{\partial} a^1 - \tilde a^2 \tilde a^3 \right) \xi \,, \\
  \delta \tilde \gamma^1 &= i \left(-2 \tilde a^1 - \tilde b^1 - 2 i \slashed{\partial} \tilde a^1 + a^2 a^3 \right) \xi \,,
 }
which implies
 \es{SUSYvarsBetter2}{
  \delta (a^1 \mp \tilde a^1) &= i \left(\gamma^1 \pm i \tilde \gamma^1 \right) \xi \,, \\
  \delta \left((b^1 - \tilde a^2 \tilde a^3) \pm (\tilde b^1 - a^2 a^3) \right) &= 
   \left[2 \slashed{\nabla} \left(\gamma^1 \pm i \tilde \gamma^1 \right) + i \left(\gamma^1 \pm i \tilde \gamma^1 \right)  \right]^T (i \sigma_2) \xi \,, \\
   \delta \left(\gamma^1 \pm i \tilde \gamma^1 \right)  &= \left(-2 (a^1 \mp \tilde a^1) - 2 i \slashed{\partial} (a^1 \mp \tilde a^1)
    +  (b^1 - \tilde a^2 \tilde a^3) \pm (\tilde b^1 - a^2 a^3) \right) \xi \,.
 }
We again see that the SUSY transformations split the integration constants into the same two sets as above, corresponding to the upper and lower signs.

The SUSY transformations given above are for an $OSp(1 | 2)$ subgroup of $OSp(2 | 2)$.  To find the ${\cal N} = 2$ transformations we should consider two independent SUSY parameters $\zeta_i$ (or $\xi_i$), with $i=1, 2$, and consider a second set of bulk gauginos that were ignored in the ${\cal N} = 1$ supergravity theory.  The coefficients $\gamma^\alpha$ and $\tilde \gamma^\alpha$ therefore become $\gamma^\alpha_i$ and $\tilde\gamma^\alpha_i$, where the extra index $i$ labels which bulk gaugino they correspond to.

\section{Global SUSY invariance of Chern-Simons-matter theories}
\label{SUSY3D}

In this Appendix we review the SUSY properties of $U(1)$ Chern-Simons theory on $S^3$ coupled to a chiral multiplet with electric charge $q = +1$.
These properties are discussed in detail in \cite{Jafferis:2010un}, but we repeat them in our conventions, because details are needed to establish the relation between the boundary behavior of bulk supergravity theory and the dual perturbed ABJM theory.

The theory on $S^3$ is obtained from the globally supersymmetric $\cn =1, D=4$ theory  in Lorentzian signature,  by analytic continuation to flat Euclidean $\R^4$ (as was done in Appendix~\ref{EUCLIDEANSUSY}),  followed by dimensional reduction to $\R^3$.  We then construct the theory on $S^3$ by inserting the corrections necessary for SUSY when the constant spinors $\e, \tilde \e$ of the $\R^3$ theory are replaced by the left-invariant Killing spinors on $S^3$, which satisfy
\be \lab{KspS3}
\nabla_i \e = \frac{i}{2a} \s_i \e\,,  \qquad \nabla_i \tilde \e =  \frac{i}{2a} \s_i \tilde \e \,.
\ee
The dimensional reduction procedure defines the matter action with its coupling
to the gauge multiplet.  The Chern-Simons Lagrangian, which describes the free dynamics of the gauge multiplet, is intrinsically three-dimensional.  We find its form 
by requiring SUSY invariance under the transformation rules obtained by the procedure above.

We start with the four-dimensional actions and Euclidean transformation rules presented in Appendix~\ref{EUCLIDEANSUSY}.  We dimensionally reduce to 3d along the 4th direction by dropping the dependence of all the fields and SUSY parameters on $x^4$.  In performing the dimensional reduction, we denote the fourth component of the gauge field $A_4 = \sigma$.  Note that our conventions \eqref{CovDers} on the gauge covariant derivatives are opposite to those of \cite{Jafferis:2010un}.

Our results for the transformation rules (including the sphere corrections) are
 \es{VectorTransf}{
  \delta A_i &=  -\frac i2  \epsilon^T (i \sigma_2) \sigma_i \tilde \lambda  -\frac i2  \tilde \epsilon^T (i \sigma_2) \sigma_i \lambda \,, \\
  \delta \sigma &= - \frac 12\epsilon^T (i \sigma_2) \tilde \lambda + \frac 12 \tilde \epsilon^T (i \sigma_2) \lambda \,, \\
  \delta D &=  -\frac 12 \epsilon^T (i \sigma_2) \left( \sigma^i \nabla_i \tilde \lambda - \frac{i}{2a} \tilde \lambda \right) 
   + \frac 12 \tilde \epsilon^T (i \sigma_2) \left(\sigma^i \nabla_i \lambda - \frac{i}{2a} \lambda \right) \,, \\
  \delta \lambda &= \left(\frac 12 \sigma^{ij} F_{ij} + i \sigma^i \partial_i \sigma + i D - \frac 1a \sigma \right) \epsilon  \,, \\
  \delta \tilde \lambda &= \left(\frac 12 \sigma^{ij} F_{ij} - i \sigma^i \partial_i \sigma - i D + \frac 1a \sigma \right) \tilde \epsilon 
 }
for a $U(1)$ vector multiplet, and
 \es{ChiralTransf}{
  \delta Z &= i \epsilon^T (i \sigma_2) \chi \,, \\
  \delta \tilde Z &= i \tilde \epsilon^T (i \sigma_2) \tilde \chi \,, \\
  \delta F &= \tilde \epsilon^T (i \sigma_2) \left(i \sigma^i D_i \chi + i \sigma \chi - \tilde \lambda Z \right) \,,\\
  \delta \tilde F &= \epsilon^T (i \sigma_2) \left(i \sigma^i D_i \tilde \chi + i \sigma \tilde \chi + \lambda \tilde Z \right) \,, \\
  \delta \chi &= F \epsilon + \left(\sigma^i D_i Z - \sigma Z + \frac{i}{2a} Z \right) \tilde \epsilon \,, \\
  \delta \tilde \chi &= \left(\sigma^i D_i \tilde Z - \sigma \tilde Z + \frac{i}{2a} \tilde Z \right) \epsilon + \tilde F \tilde \epsilon 
 }
for a chiral multiplet.  Here, $D_i \chi = (\nabla_i - i A_i) \chi$, $D_i Z = (\partial_i - i A_i) Z$, etc.  When $a = \infty$, these expressions are the dimensional reduction of \eqref{Gauge4dTransf} and \eqref{ChiralTransf4d}.  The 3d actions invariant under these transformation rules are
  \es{3dActions}{
   S_{\text CS} &= \frac{ik}{4 \pi} \int d^3x \, \big[\e^{ijk} A_i \pa_j A_k  - \sqrt{g}(\l^\dagger \l + 2 i \s D)\big] \,, \\
  S_\text{chiral} &= \int d^3 x\, \sqrt{g} \Biggl(D^i \tilde Z D_i Z + \sigma^2 \tilde Z Z + i \tilde \chi^T (i \sigma_2) \sigma^i D_i \chi 
   + i \tilde \chi^T (i \sigma_2) \sigma \chi  - \tilde F F \\
   &\qquad\qquad\qquad\qquad\qquad+ \lambda^T (i \sigma_2) \tilde Z \chi + \tilde \chi^T (i \sigma_2) Z \tilde \lambda - D \tilde Z Z + \frac{3}{4a^2} \tilde Z Z \Biggr) \,.
 }
When $a = \infty$, the matter action $S_\text{chiral}$ is obtained by dimensional reduction from its 4d counterpart \eqref{Chiral4dEuc}.
 
It is straightforward, but tedious, to check how the SUSY algebra is realized on the chiral multiplet.  Denoting by $\delta$ and $\tilde \delta$ the contributions to the variations \eqref{VectorTransf}--\eqref{ChiralTransf} that are proportional to $\epsilon$ and $\tilde \epsilon$, respectively, we have
 \es{SUSYCommutators}{
  [\delta, \tilde \delta] Z = \left[-i \epsilon^T (i \sigma_2) \sigma^i \tilde \epsilon D_i + \epsilon^T (i \sigma_2) \tilde \epsilon
   \left( i \sigma + \frac{1}{2a} \right) \right] Z \,, \\
 [\delta, \tilde \delta] \tilde Z = \left[-i \epsilon^T (i \sigma_2) \sigma^i \tilde \epsilon D_i - \epsilon^T (i \sigma_2) \tilde \epsilon
   \left( i \sigma + \frac{1}{2a} \right) \right] \tilde Z \,, \\
   [\delta, \tilde \delta] \chi = \left[-i \epsilon^T (i \sigma_2) \sigma^i \tilde \epsilon D_i + \epsilon^T (i \sigma_2) \tilde \epsilon
   \left( i \sigma - \frac{1}{2a} \right) \right] \chi \,, \\
 [\delta, \tilde \delta] \tilde \chi = \left[-i \epsilon^T (i \sigma_2) \sigma^i \tilde \epsilon D_i - \epsilon^T (i \sigma_2) \tilde \epsilon
   \left( i \sigma - \frac{1}{2a} \right) \right] \tilde \chi \,, \\ 
   [\delta, \tilde \delta] F = \left[-i \epsilon^T (i \sigma_2) \sigma^i \tilde \epsilon D_i + \epsilon^T (i \sigma_2) \tilde \epsilon
   \left( i \sigma - \frac{3}{2a} \right) \right] F \,, \\
 [\delta, \tilde \delta] \tilde F = \left[-i \epsilon^T (i \sigma_2) \sigma^i \tilde \epsilon D_i - \epsilon^T (i \sigma_2) \tilde \epsilon
   \left( i \sigma - \frac{3}{2a} \right) \right] \tilde F \,.   
 }
Writing the supersymmetry variations in terms of the supercharges:
 \es{deltaToQ}{
  \delta = i \epsilon^T (i \sigma_2) Q \,, \qquad
   \tilde \delta = i \tilde \epsilon^T (i \sigma_2) \tilde Q \,,
 } 
it is not hard to see that $[\delta, \tilde \delta] = \epsilon^T(i\sigma_2) \{Q, \tilde Q^T (i \sigma_2)\} \tilde \epsilon$, so the expressions \eqref{SUSYAlgebra} are consistent with the supersymmetry algebra
 \es{SUSYAlgebra}{
  \{ Q, \tilde Q^T i \sigma_2 \} = \sigma^i J_i + i q \sigma + \frac{1}{a} R \,,
 }
where $J_i$ is an $SU(2)_r$ isometry, $q$ is the gauge charge, and $R$ is the R-charge.  For the free chiral multiplet $(Z, \chi, F)$, the gauge charge is $+1$ and the R-charges are $(1/2,  -1/2,  -3/2)$.  The anti-chiral multiplet $(\tilde Z, \tilde \chi, \tilde F)$ has opposite gauge and R-charges.

The anticommutators in \eqref{SUSYAlgebra}, as well as the entire discussion in this Appendix, involves the odd generators of $OSp(2|2)_r$, whose bosonic part contains the $SU(2)_r$ subgroup of the $SO(4) \cong SU(2)_\ell \times SU(2)_r$ isometry of $S^3$.  One can repeat this discussion by using $OSp(2|2)_\ell \supset SU(2)_\ell$ to couple the flat space theory to curvature, which amounts to sending $a \to -a$ in all the above formulas.  In particular, the Killing spinors \eqref{KspS3} would now be right-invariant.

%%%%%%%%%%%
\bibliographystyle{ssg}
\bibliography{massive}

\end{document}